\documentclass{article}
\usepackage{amsmath}
\usepackage{authblk}
\usepackage[pdftex,breaklinks,colorlinks=true,urlcolor=blue,linkcolor=blue, citecolor=blue]{hyperref}
\hypersetup{linktocpage} 
\usepackage{geometry}
\usepackage[backend=biber, sorting=none, style=ieee, citestyle=numeric-comp]{biblatex}
\usepackage{amsfonts}
\usepackage{amssymb}
\usepackage{mathtools}
\usepackage{mathrsfs}
\usepackage{tikz}
\usepackage{float}

\usepackage{multirow}

\usepackage{xcolor}

\numberwithin{equation}{section}	

\newcommand{\blank}[1]{}

\newcommand{\poletype}[3]{\text{Z-{#2}efect{#3}}}
\newcommand{\zerotype}[3]{\text{P-{#2}efect{#3}}}
\newcommand\plusdot{\stackrel{\mathclap{\normalfont\mbox{.}}}{+}}
\newcommand{\fp}{\, .}
\newcommand{\ca}{\, ,}
\newcommand{\tr}[2]{\left\langle {#1}, {#2} \right\rangle}
\newcommand{\res}[1]{\text{Res}_{q}\left({#1}\right)}
\newcommand{\prof}{|^{\mathfrak{f}}}
\newcommand{\proh}{|^{\mathfrak{h}}}

\newcommand{\lax}[1]{\mathcal{{#1}}}
\newcommand{\remsph}{\mathbb{C} P^{1}}
\newcommand{\nn}{\nonumber}
\newcommand{\pp}{(\textbf{\textit{q}})}
\newcommand{\sigpi}{\Sigma_{q}}
\newcommand{\dsig}{d_{\Sigma}}
\newcommand{\man}{\Sigma \times \mathbb{C} P^{1}}
\newcommand{\cdel}{\widebar{\partial}}
\newcommand{\p}{\partial}

\makeatletter
\newcommand*\rel@kern[1]{\kern#1\dimexpr\macc@kerna}
\newcommand*\widebar[1]{%
	\begingroup
	\def\mathaccent##1##2{%
		\rel@kern{0.8}%
		\overline{\rel@kern{-0.8}\macc@nucleus\rel@kern{0.2}}%
		\rel@kern{-0.2}%
	}%
	\macc@depth\@ne
	\let\math@bgroup\@empty \let\math@egroup\macc@set@skewchar
	\mathsurround\z@ \frozen@everymath{\mathgroup\macc@group\relax}%
	\macc@set@skewchar\relax
	\let\mathaccentV\macc@nested@a
	\macc@nested@a\relax111{#1}
	\endgroup
}

\pgfdeclareimage{Chern-Simons-holonomy}{Chern-Simons-holonomy}
\pgfdeclareimage{WZW-CP}{WZW-CP}
\pgfdeclareimage{many-poles}{many-poles}

\title{Four-Dimensional Chern-Simons and Gauged Sigma Models}
\author{Jake Stedman\thanks{jake.williams@kcl.ac.uk}}
\affil{Department of Mathematics, King's College London, \protect\\
	 Strand, London, WC2R 2LS, UK}

\begin{document}
	
	\begin{flushright}  {~} \\[-12mm]
		{KCL-MTH-21-01}
	\end{flushright}
	
	\thispagestyle{empty}
	
	\nopagebreak
	
	\begingroup
		\let\newpage\relax%
		\maketitle
	\endgroup

	\begin{abstract} 
        In this paper, we introduce a new method for constructing gauged $\sigma$-models from four-dimensional Chern-Simons (4d CS) gauge theory. 
        We begin with a review of recent work by several authors on the classical generation of integrable $\sigma$-models from 4d CS. 
        In this approach, a gauge field is required to satisfy certain boundary conditions on two-dimensional defects inserted into the bulk. 
        Using these boundary conditions, the equations of motion are solved, and the result is substituted back into the action. 
        This yields a $\sigma$-model whose integrability is guaranteed because the 4d CS field is gauge equivalent to a Lax connection. 
        
        Using a theory consisting of two 4d CS fields coupled together on new classes of ``gauged'' defects, we construct gauged $\sigma$-models and identify a unifying action. 
        These models are conjectured to be integrable because the 4d CS fields remain gauge equivalent to two Lax connections. 
        Finally, we consider two examples: the gauged Wess-Zumino-Witten model and the nilpotent gauged Wess-Zumino-Witten models. 
        This latter model is of note as one can find the conformal Toda models from it.
    
	\end{abstract}

	\newpage
	
		\tableofcontents
	
	\newpage
	
	\section{Introduction}

	Over the last two decades, several groups have asked whether gauge theories can be used to identify properties of conformal field theories (CFTs), vertex operator algebras (VOAs), and integrable models (IMs). 
    We know of three such examples. 
    The first is by Fuchs et al., who have used topological quantum field theories to analyse CFTs \cite{Fuchs:2002cm,Fuchs:2003id,Fuchs:2004dz,Fuchs:2004xi,Fjelstad:2005ua}; and the second, by Beem et al., has shown a deep connection between 4d $\mathcal{N}=2$ superconformal field theories and VOAs \cite{Beem:2013sza,Beem:2017ooy}. 
    Our paper is interested in the final example, whose start lies in Costello's work in \cite{Costello:2013zra,Costello:2013sla}, and which he subsequently expanded upon in collaboration with Witten and Yamazaki in \cite{Costello:2017dso,Costello:2018gyb,Costello:2019tri}.
    In this series of papers, the authors introduce a new gauge theory, called \textit{four-dimensional Chern-Simons} (4d CS), and use it to explain several properties of two-dimensional IMs. 
    For example, in \cite{Costello:2017dso,Costello:2018gyb} the Wilson lines of 4d CS are used to find the $R$-matrix and Quantum group structures of lattice and particular scattering models, while in a fourth paper \cite{Costello:2021zcl}, 't-Hooft and $Q$-operators are shown to be related.

	Our focus is on the third paper \cite{Costello:2019tri}, in which it was proven that classical 4d CS reduces to a two-dimensional integrable $\sigma$-model when suitably chosen defects (defined by a set of ``boundary conditions'') are introduced. 
    To perform this reduction, one uses the boundary conditions to solve the 4d CS equations of motion in terms of a group element $\hat{g}$, whose values on the defects are the $\sigma$-model's fields. 
    This is analogous to the construction of the Wess-Zumino-Witten (WZW) model from three-dimensional Chern-Simons (3d CS) given in \cite{Elitzur:1989nr}, although there the model sits on the boundary rather than on defects.
    
    From a physical point of view, this work is particularly exciting because $\sigma$-models exhibit many of the phenomena present in non-abelian gauge theories, such as confinement, instantons or anomalies \cite{DAdda:1978dle,Witten:1978bc,DAdda:1978vbw,Abdalla:1982yd}; and their integrability ensures the existence of exact solutions, see e.g. \cite{Abdalla:1986xb,Abdalla:1984gm,Faddeev:1985qu,Eichenherr:1981sk}.
	

	Alongside these developments, Vicedo, in \cite{Vicedo:2019dej}, observed that the gauge field $A$ of 4d CS theory is gauge equivalent to the Lax connection $\lax{A}$ (with or without a spectral parameter) of a two-dimensional integrable $\sigma$-model (further details concerning Lax connections and their origin within 4d CS will be discussed below).
    This work was expanded upon in \cite{Delduc:2019whp} by Delduc, Lacroix, Magro and Vicedo (DLMV), who construct a general action for ``genus zero'' integrable $\sigma$-models called the unified $\sigma$-model action. 
    This result is remarkable for two reasons: firstly, it produces a geometric method for constructing an integrable $\sigma$-model's Lax connection; and secondly, it gives a general action from which a class of $\sigma$-models can be found - provided their Lax connection is known. 
    We will refer to this construction as the \textit{DLMV construction} throughout the following.

    In this paper we ask: can one use 4d CS to construct gauged $\sigma$-models whose target spaces are cosets (manifolds of the form $G/H$ where $G$ and $H \subseteq G$ are groups)? 
    We have two motivations for asking this question: the first being that gauged $\sigma$-models include the GKO constructions \cite{Goddard:1984hg,Goddard:1984vk,Goddard:1986ee} from which one can, possibly, find all rational conformal field theories (RCFTs). 
    These theories are integrable and occasionally have a known Lax connection. 
    It is thus expected that they to fall under the 4d CS umbrella. 
    Our second motivation is the existence of a few special cases, such as symmetric space models in \cite{Costello:2019tri} or $\lambda$- and $\eta$-deformations (which can realise coset models for certain subgroups of $H \subset G$) \cite{Delduc:2019whp,Schmidtt:2019otc,Tian:2020ryu}. 
    Unfortunately, these do not offer up much of a guide as to a more general method.
    
    Our primary result is a proof that one can generate gauged $\sigma$-models by reducing a new theory, which we call \textit{doubled four-dimensional Chern-Simons} (doubled 4d CS). 
    This is constructed from two copies of standard 4d CS, with the gauge groups (resp. fields) $G$ and $H \subseteq G$ (resp. $A$ and $B$), that are then coupled together on a new class of two-dimensional defect, collectively called \textit{gauged defects}. 
    This result is analogous to the work of Moore and Seiberg in \cite{Moore:1989yh}, where it was shown that the GKO constructions are the boundary theory of a doubled 3d CS model - see also \cite{Gawedzki:2001ye}.
	
	As before, $A$ and $B$ are gauge equivalent to a pair of Lax connections $\lax{A}$ and $\lax{B}$ (with or without spectral parameter). 
    We find that the integrable $\sigma$-model's equations of motion are the Lax equations of this pair, together with a set of gauging constraints which follow from the boundary conditions on the defects.
    Thus, we conjecture that these models are integrable. 
	By following arguments similar to those of Delduc et al. in \cite{Delduc:2019whp}, we find a unified gauged $\sigma$-model from which a large class of integrable gauged $\sigma$-models can be found. 
    A notable feature of these theories is that their target space is a homogeneous reductive space $G/H$ (defined below), which is, in general, less restricted than a symmetric space. 
    To our knowledge, no previous examples of such models were known, outside of the special case already mentioned.

    A reason for anticipating this construction comes from the gauged WZW model. 
    In particular, it is known that this model can be found from the difference of two WZW models (see appendix \ref{WZW convetions}), each of which can be found from 4d CS theory. 
    Alternatively, a similar argument can be made via 3d CS, which we know is $T$-dual to the 4d theory \cite{Yamazaki:2019prm}. 
    Using this fact, one would expect doubled 4d CS to be $T$-dual to doubled 3d CS. 
    Since the gauged WZW model is the boundary theory of the latter, one would hope to find it from the former.

	The structure of this paper is as follows: in section \ref{4dcs section} we define 4d CS theory, deriving its equations of motion and boundary conditions amongst other properties. In section \ref{section: Integrable Sigma Models} we review the construction of integrable $\sigma$-models by Delduc et al. in 4d CS. 
    In section \ref{doubled 4dcs section}, we define the doubled 4d CS theory, its associated auged defects and describe its gauge invariance. 
	
	In section \ref{section:UGSM}, we use the DLMV approach to derive the unified gauged $\sigma$-model and construct the normal gauged WZW model and a nilpotent gauged WZW model. 
    These examples are notable for two reasons: the first is that the normal gauged WZW model gives an action for the GKO constructions as described in \cite{Karabali:1989dk,Karabali:1988au,Hwang:1993nc,Gawedzki:1988hq,Gawedzki:1988nj}; the second reason is that the Toda field theories can be found from both of these actions. 
    In the former case, this is a quantum equivalence with the $G_{k} \times G_{1} /G_{k+1}$ GKO model, as shown in \cite{Eguchi:1989hs}, while in the latter case this is proven via a Hamiltonian reduction as shown in \cite{Balog:1990mu}. 
    It was also shown in \cite{Balog:1990mu} that one can find the $W$-algebras from the nilpotent gauged WZW model. 
	
	In section \ref{conclusion} we summarise our results and comment on a few potential directions of this research.

    \section*{Acknowledgements}

	I would like to thank my supervisor G\'{e}rard Watts for proposing this problem and the support he has provided during our many discussions. I would also like to thank Ellie Harris and Rishi Mouland for our discussions; Nadav Drukker who kindly provided comments on a previous version of this manuscript; Benoit Vicedo for his comments; and finally the anonymous referees whose comments we feel have greatly improved the following work.  This work was funded by the STFC grant ST/T506187/1.
	
\section{The Four-Dimensional Chern-Simons Theory}

\label{4dcs section}

	In this section, we start by introducing the 4d CS action and derive the equations of motion. 
    Afterwards, we define the various defect boundary conditions used to construct IMs later on, and we conclude by describing the action's gauge invariance.
	
	\subsection{The Action}

    Four-dimensional Chern-Simons is a gauge theory constructed from a connection $A = A_\mu d x^\mu$ on a principal bundle over a spacetime of the form $M = \Sigma \times C$, where $\Sigma$ and $C$ are both two-dimensional surfaces. 
    Further still, $C$ is assumed to be a complex manifold equipped with a meromorphic one-form $\omega$ that has a set of zeros $Z$ and poles $P$. 
    Its gauge group $G$ is complex semi-simple.

    To construct the 4d CS action, we need one final structure. 
    This is a non-degenerate symmetric bilinear form $\tr{\cdot}{\cdot}$ proportional to the Killing form of the complex Lie algebra $\mathfrak{g}$ for the Lie group $G$. 
    We shall take $A_\mu = A_\mu^a T^a$ to be in the adjoint representation\footnote{One should note that this is only possible when the adjoint representation is non-trivial. 
    If the adjoint representation is degenerate, such as for $U(1)$, then one must use an alternative representation.} of $\mathfrak{g}$, and normalise the basis elements $T^a$ such that $\tr{T^{a}}{T^{b}} = \delta^{ab}$. 
    Thus, the 4d CS action is: 
    \begin{equation}
		S(A) = \frac{i}{8 \pi^2} \int_M \omega \wedge \tr{A}{dA + \tfrac{1}{3} [A, A ]} \label{cwy action} \ca
	\end{equation} 
    where $[A,A] = [A_\mu, A_\nu] dx^\mu \wedge dx^\nu$. 
    Note, when both entries in $\tr{\cdot}{\cdot}$ are differential forms, there is an implicit wedge product between the two entries. 
    As mentioned in the introduction, we will discuss two classes of the 4d CS action in the following and refer to \eqref{cwy action} as the standard action, or theory, for short. 

    The IMs one can generate using 4d CS depend upon the choices one makes when fixing $G$, $C$ and $\omega$. 
    {For simplicity's sake, we fix $C$ to be the Riemann sphere, $ \remsph$, and therefore note that the integrable theories we construct belong to the ``genus zero'' class.
    When working with a set of complex coordinates $z$ and $\bar{z}$ on $\mathbb{C} P^1$, this restriction implies that $\omega = \varphi(z) dz$, where $\varphi(z)$ is a rational function of $z$ called the \textit{twist function}. 
    If we impose as a further restriction that $\omega$ has at most double poles, which we will throughout, and denote by $P_\mathrm{fin}$ the subset of $P$ not including infinity, then from a partial fraction expansion we have: 
    \begin{equation}
		\omega = k_{0}^{(-1)} dz + \sum_{q \in P_\mathrm{fin}} \sum_{l=0}^{n_q-1} \frac{k_{q}^{(l)} \, dz}{(z-q)^{l+1}} \ca 
		\label{pf omega}
	\end{equation}
    where\footnote{$k_{0}^{(-1)} \neq k_0^{-1}$.}: 
    \begin{equation}
        k_{q}^{(l)} = \res{(z-q)^{l}\varphi(z)} \qquad \text{and} \qquad k_0^{(-1)} = \text{Res}_{0} \, (\varphi(z)/z)
    \end{equation}
    and with $n_q$ denoting the order of the pole $q$. 
    We similarly, denote the degree of a zero $\zeta$ of $\omega$ by $m_\zeta$.} 

    {If we count with multiplicity the number $N_Z$ of zeros and $N_P$ of poles of $\omega$ we find: 
    \begin{equation}
        N_P = \sum_{q\in P} n_q \ca \qquad N_Z = \sum_{\zeta\in Z} m_\zeta \;,    
    \end{equation}
    which, by the Riemann-Roch theorem, must satisfy $N_P = N_Z +2$. 
    Since $N_Z \geq 0$ it immediately follows that $\omega$ has at least one pole, which we can always place at $z=\infty$ by the transformation $z \rightarrow (1/z)+q$. 
    This is possible because the isometry group of the Riemann sphere is the set of M{\"o}bius transformations, within which $z \rightarrow (1/z)+q$ is contained (since inversions and translations are). 
    From here on, we assume $\omega$ has a pole at infinity and that $P = P^\textrm{fin} \cup \{\infty\}$.}

    The presence of $\omega$'s poles and zeros requires us to carefully consider of the behaviour of $A$ near these locations. 
    We will call the resulting restriction on $A$ ``boundary conditions''. 
    Since a point $(q,\bar{q})$ in $\remsph$ corresponds to a two-dimensional surface $\Sigma_q := \Sigma \times (q, \bar{q})$ embedded within $M$, these boundary conditions define two-dimensional defects in the theory. 
    The defects associated with zeros are called \poletype{t}{d}{s} and those associated with poles, \zerotype{t}{d}{s}. 
    We discuss these in sections \ref{section:Lax connection} and \ref{boundary conditions in cwy} respectively.
	

	While $\Sigma$ can be more general, we will usually take it to be $\mathbb R^2$.
	Generically, the theory does not depend on a metric on $\Sigma$, but we find it helpful to write the coordinates on $\Sigma$ as $x^\pm$ as the result of the construction is a two-dimensional Lorentzian theory in which $x^\pm$ are ``light-cone coordinates''.
    We will see that $\Sigma$ is the worldsheet of our $\sigma$-models.
	
	{To aid the clarity of the following, we emphasise two important facts. 
    The first of these being that \eqref{cwy action} is not invariant under all diffeomorphisms of $M$, or in the vernacular ``topological'', but rather ``semi-topological'', \textit{i.e.} invariant under all diffeomorphisms of $\Sigma$. 
    Na{\"i}vely, one would expect the former as the action is constructed from wedge products, and contains no metric. 
    However, it is instead the latter because $\varphi(z)$ does not transform as a vector, unlike $A_\mu$.}
	
	Our second fact is that \eqref{cwy action} has an additional \textit{unusual gauge invariance} of the form:
    \begin{equation}
		A_{z} \rightarrow A_{z} + \chi_{z} \ca \label{unusual gauge trans}
	\end{equation}
	where the generator $\chi_{z}$ can be any $\mathfrak{g}$ valued function. 
    To see this, we note that the $z$ components of the gauge field and the exterior derivative $d$, $A_{z}dz$ and $dz\partial_{z}$, both fall out of \eqref{cwy action} due to the wedge product with $\omega = \varphi(z)dz$ and the condition $dz \wedge dz = 0$. 
    Due to \eqref{unusual gauge trans}, all field configurations of $A_{z}$ are gauge equivalent and we are thus free to set $A_{z} = 0$. 
    As a result, the gauge field $A$ will be given by:
	\begin{equation}
		A = A_{\Sigma} + \widebar{A} \ca \label{field expansion}
	\end{equation}
    through the following and in which $\bar A = A_{\bar z}d\bar z$ and $A_{\Sigma}=A_+dx^+ + A_-dx^-$ (the restriction of $A$ to $\Sigma$).
	
	For the same reason, we also drop the $dz$ term from the exterior derivative $d = \dsig + \cdel + \partial$, where $\dsig$ is the exterior derivative on $\Sigma$ and $\cdel = d \bar{z} \, \p_z$ and $\partial = d \bar{z} \partial_{\bar{z}}$ are the Dolbeault operators on $\mathbb{C} P^1$. 
    By an abuse of notation, we will also write the resulting modified exterior derivative as $d$, where:
	\begin{equation}
		d = \dsig + \cdel \;. \label{eq: exdiv expansion}
	\end{equation} 
    Notice that even though a gauge transformation does not contain a wedge product with $dz$, we are still free to do this because we can always set $A_z$ to zero, as has just been described.

    Finally, let us briefly comment on the regularity of the action. 
    Our motive for doing so is that to produce $\sigma$-models from the field configurations constructed below, we need the action to be finite. 
    This is not generically the case if $\omega$ has more than simple poles. 
    To demonstrate why this might be the case, suppose, without loss of generality, that there exists a double pole at $z=0$ and consider the action in the disc around the pole:
    \begin{align}
        S_U(A) &= \frac{i}{8 \pi^2} \int_{\Sigma \times U} \frac{dz}{z^2} \wedge \tr{A}{dA + \tfrac{1}{3} [A,A]} \nn \\
        &= \frac{i}{8 \pi^2} \int_{\Sigma \times U} \frac{dz\wedge d \bar{z}}{z^2} \wedge \left( 2 \tr{A_{\bar{z}}}{ \dsig A_{\Sigma} + \tfrac{1}{2}[A_{\Sigma},A_{\Sigma}]} - \tr{A_{\Sigma}}{\p_{\bar{z}} A_{\Sigma}} \right) \ca
    \end{align}
    where in the second equality we have expanded the gauge field into its components $A = A_{\bar{z}} d \bar{z} + A_{\Sigma}$, integrated by parts a derivative in $d_{\Sigma}$ and sent a total derivative to zero.
    After performing the change of coordinates, $z = r e^{i \theta}$, this integral becomes:
    \begin{equation}
        S_U(A) = \frac{1}{4 \pi^2} \int_{\Sigma \times U} e^{-2i \theta} \frac{dr \wedge d \theta}{r} \wedge \left( \tr{A_{\bar{z}}}{\dsig A_{\Sigma} + \tfrac{1}{2}[A_{\Sigma},A_{\Sigma}]} - \tr{A_{\Sigma}}{ \p_{\bar{z}} A_{\Sigma}} \right) \ca \label{eq: singular term}
    \end{equation}
    which is clearly singular unless $A_{\Sigma}$ goes to zero at $r=0$ at least as $O(r)$. 
    We shall see that this is indeed the case for the boundary conditions discussed below. 

    The approach we are using, which appears implicitly in \cite{Costello:2019tri}, is a restriction on the bundles we consider to those satisfying boundary conditions such that there are no singular terms. 
    However, for the sake of completeness, we observe that a second method exists in which any singular terms are subtracted from the action, see \cite{Benini:2020skc} for further details.

	\subsection{The Equations of Motion and Defect Boundary Conditions}

    \label{section: 4dcs EOM}
		
	Our analysis is entirely classical and focuses on solving the standard action's equations of motion. As usual, these are easily found from the demand that the following variation vanish:
	\begin{equation}
		\delta S_{\text{4dCS}}(A) = \frac{i}{4 \pi^2} \int_{M} \omega \wedge \tr{ F(A)}{\delta A} - \frac{i}{8 \pi^2}\int_{M} \omega \wedge \cdel \tr{A}{\delta A} \ca \label{eq:Svar}
	\end{equation} 
    where $F = dA + \tfrac{1}{2}[A,A]$. 
    We treat the two terms on the right-hand side separately, by demanding that they vanish independently.
    We do this because the second term simplifies to a sum over terms evaluated at the poles of $\omega$, as we demonstrate momentarily.

    By demanding that the first term vanishes, we find the \textit{bulk equations of motion}:
    \begin{equation}
		\omega  \wedge (\cdel A_{\Sigma} + \dsig \widebar{A} + \widebar{A} \wedge A_{\Sigma} + A_{\Sigma} \wedge \widebar{A} ) =0 \ca \qquad d_{\Sigma} A_{\Sigma} + A_{\Sigma} \wedge A_{\Sigma} = 0 \, .
		\label{cwy eom} 
	\end{equation}
    Importantly, the field configurations which satisfy the above equations are allowed to be singular at the points in $Z$. 
    We will see shortly that this is because $\omega$ has zeros at these points.
    These singularities are the \poletype{t}{d}{s} defined in the subsection next but one. 

    The second term on the right-hand side of \eqref{eq:Svar} collapses to a sum over the poles of $\omega$. 
    Since similar calculations will appear further down, we will show this for the more general expression\footnote{In the second of these equalities we have integrated by parts and sent the resulting total derivative to zero since the Riemann sphere is compact.}: 
    \begin{equation}
        I = \frac{i}{8 \pi^2}  \int_{M} \omega \wedge \cdel \xi = - \frac{i}{8 \pi^2} \int_M \cdel \omega \wedge \xi \ca \label{eq: boundary terms}
    \end{equation} 
    in which $\xi$ is a two-form on $\Sigma$, and then apply the result to \eqref{eq:Svar}. 
    To evaluate this integral, we first cover the Riemann sphere with two charts, $N$ and $S$, whose respective coordinates are $z$ and $w=z^{-1}$. 
    These charts are chosen to ensure that the poles of $\omega$ contained within the subset $P_{\text{fin}}$ occur within $N$, while $S$ only contains the pole at $\infty$. 
    Using this atlas, we can decompose the integral into $I = I_N + I_S$ (where the subscript indicates the restriction to the respective chart) and calculate each term separately.

    Starting with $I_N$, we substitute in the expression for $\omega$ given in equation \eqref{pf omega}, and note that any terms holomorphic in $z$ are annihilated by $\cdel$ holomorphic functions. 
    We thus find:
    \begin{align}
        I_N &= - \frac{i}{8 \pi^2} \sum_{q \in P_{\text{fin}}} \sum_{l=0}^{n_q-1} k_{q}^{(l)} \int_{\Sigma \times N} \cdel(z-q)^{-l-1} \wedge dz \wedge \xi \nn \\
        &= - \frac{i}{8 \pi^2} \sum_{q \in P_{\text{fin}}} \sum_{l=0}^{n_q-1} (-1)^l \frac{k_{q}^{(l)}}{l!} \int_{\Sigma \times N}  \cdel \, \partial_z^l (z-q)^{-1} \wedge dz \wedge\xi \nn \\
        &= \frac{1}{4 \pi} \sum_{q \in P_{\text{fin}}} \sum_{l=0}^{n_q-1} (-1)^l \frac{k_{q}^{(l)}}{l!} \int_{\Sigma \times N} d \bar{z} \wedge dz \, \partial_z^l \, \delta^2 (z-q) \wedge \xi \nn \\
        &= \frac{1}{4 \pi} \sum_{q \in P_{\text{fin}}} \sum_{l=0}^{n_q-1} \frac{k_{q}^{(l)}}{l!} \int_{\Sigma_q} \partial_z^l  \xi \ca 
    \end{align}
    where in the second equality, we have used $(z-q)^{-l-1} = (-l)^l \partial_z^l (z-q)^{-1}/l! \, $, in the third $\partial_{\bar{z}} (z-q)^{-1} = \delta^2 (z-q)$ and in the final one evaluated the integral over $N$. 
    In the final expression above $\Sigma_q$ indicates an integral over $\Sigma$ with the integrand evaluated at $(z,\bar{z}) = \textbf{\textit{q}} := (q, \bar{q})$.

    Within $S$ we have $\omega = \widebar{\varphi}(w) dw$, where $\widebar{\varphi}(w) =  -\varphi(1/w) /w^2$, and a similar calculation to that just performed yields: 
    \begin{equation}
        I_S = \frac{1}{4 \pi} \sum_{l = 0}^{n_\infty -1} \frac{k_{\infty}^{(l)}}{l!} \int_{\Sigma_q} \partial_w^l \xi \fp
    \end{equation}
    Thus, we find:
    \begin{equation}
        I =  \frac{1}{4 \pi} \sum_{q \in P} \sum_{l=0}^{n_q-1} \frac{k_{q}^{(l)}}{l!} \int_{\Sigma_q} \partial_z^l \xi \label{integral over poles} \ca
    \end{equation} 
    where we treat $z$ as the local coordinate of the pole in this expression. 
    Throughout the following, it is often more convenient to write and evaluate \eqref{integral over poles} as a sum over residues. 
    This is possible due to the following calculation. 
    Let $V_{q} \subset \mathbb{C} P^{1}$ denote an open region which contains the pole $q$ only, thus we have:
	\begin{align}
    	\int_{\sigpi} \res{\omega \wedge \xi} &= \int_{\Sigma \times V_{q}} d\bar{z} \wedge \delta^2(z-q) \partial_{z}^{n_q-1} \left( \frac{(z-q)^{n_q}}{(n_q-1)!} \omega \wedge \xi \right) \nn \\ 
		 &= \sum_{l=0}^{n_q-1} \int_{\Sigma \times V_{q}} d\bar{z} \wedge \delta^2(z-q) \binom{n_q-1}{l} \partial_{z}^{n_q-l-1} \left( \frac{(z-q)^{n_q-l}}{(n_q-1)!} (z-q)^{l} \omega \right) \wedge \partial_{z}^{l} \xi \nn \\
		 &= \sum_{l=0}^{n_q-1} \frac{k_{q}^{(l)}}{l!} \int_{\sigpi} \partial_{z}^{l} \xi \ca \label{generic residue integral}
	\end{align}
	where in the final equality we have cancelled $(n_q-1)!$ with the same term in the binomial coefficient, used $k_{q}^{(l)} = \res{(z-q)^{l} \omega}$ and evaluated an integral over $V_{q}$.

    Using \eqref{integral over poles}, we can collapse the second term on the right-hand side of \eqref{eq:Svar} by setting $\xi = \tr{A}{\delta A}$. 
    The requirement that the resulting ``boundary terms'' vanish produces the condition: 
	\begin{equation}
		\delta S_{\partial M}(A,\delta A) = \frac{1}{4 \pi} \sum_{q \in P} \sum_{l=0}^{n_q-1} \frac{k_{q}^{(l)}}{l!} \int_{\Sigma_q} \partial_z^l \tr{A}{\delta A} = 0 \ca \label{beom 1}
	\end{equation}
	whose solutions a set of ``boundary conditions'' that $A$ satisfies at the poles of $\omega$. 
    Imposing these conditions introduces a set of two-dimensional defects spanning $\Sigma_q$ that we call \zerotype{t}{d}{s}. 

    \subsubsection{The Lax Connection}

    \label{section:Lax connection}

    In this section, we introduce the notion of a Lax connection and explain its origins within 4d CS. 

    Let $\lax{L}$ be a connection on a principal bundle over a worldsheet $\Sigma$, where the components of $\mathcal{L}$ are functions of a $\sigma$-model's fields $\{ g_q\}$. 
    $\mathcal{L}$ is said to be a Lax connection if the following properties are satisfied \cite{Babelon:2003qtg}: 
    \begin{enumerate}
		\item The equation $\dsig \lax{L} + \frac{1}{2} [\lax{L} , \lax{L}] = 0$ gives the equations of motion for the IM.
	\end{enumerate}
	It is called a Lax connection with spectral parameter if, in addition, it satisfies:
    \begin{enumerate}
		\item[2.] $\lax{L}$ has a meromorphic dependence upon on a complex parameter $z$, called the spectral parameter,
		\item[3.] One can obtain an infinite set of Poisson-commuting charges from the Taylor expansion in $z$ of the trace of (products of) the monodromy matrix of $\lax{L}$ at fixed time: 
        \begin{align}
            U(z,t) = \text{P} \exp \int_{\gamma} \lax{L}(z,t) \ca
        \end{align} 
        where $\gamma$ is a curve spanning a timeslice of $\Sigma$. 
	\end{enumerate} 
    For clarity's sake, let's illustrate the third point by fixing $\Sigma$ to be the cylinder $S^1 \times \mathbb{R}$ parametrised by the coordinates $(\theta,t)$, and with the assumption that the Lax connection is periodic, $\lax{L}(\theta = 0) = \lax{L}(\theta = 2 \pi)$. 
    In this scenario the trace of $U(z,t)$ is:
	\begin{align}
		W(z) = \text{Tr} \, \text{P} \exp \int_{0}^{2 \pi} \lax{L}_{\theta}(z,t) \, d \theta \ca
	\end{align} 
    which is independent of $t$ because:
	\begin{equation}
		\partial_{t} W(z,t) = \text{Tr} [U(z,t),\lax{L}_{t}] = 0 \fp \label{cons. first}
	\end{equation}
    By Taylor expanding $W(z)$ in $z$ it follows from \eqref{cons. first} that the coefficient of each power is conserved in time. 
    This set of coefficients is the infinite stack of charges associated with $\lax{L}$. 
    They Poisson commute with one another because $\lax{L}$ has the Maillet bracket as its Poisson bracket, see \cite{Maillet:1985ek,Maillet:1985fn}. 

    How then do Lax connections arise within 4d CS? 
    To explain this, we work in the Lax gauge: 
    \begin{equation}
        \widebar{A} = 0 \ca \qquad A_{\Sigma} = \lax{A} = \mathcal{A}_{+} dx^{+} + \mathcal{A}_{-} dx^- \ca \label{eq: lax gauge}
    \end{equation} 
    in which the bulk equations of motion reduce to:
    \begin{align}
		\dsig \lax{A} + \frac{1}{2} [\mathcal A , \mathcal A ] = 0 \ca \qquad \omega \wedge \cdel \lax{A} = 0 \fp
	\end{align} 
    The first of these equations is simply the statement that $\lax{A}$ is flat on the worldsheet $\Sigma$, and the second implies that $\lax{A}$ depends meromorphically on the spectral parameter $z$ with poles at the zeros of $\omega$ (we shall show this shortly). 
    Thus, the first two conditions of a Lax connection are satisfied. 
    What about the third? 
    This condition is also satisfied as the Wilson lines of 4d CS: 
    \begin{equation}
        W(z) = \text{Tr} \, \text{P} \exp \int_\gamma A \ca
    \end{equation} 
    are gauge invariant observables and, by of the equations of motion, generate an infinite stack of conserved charges in the way we've just described. 
    That these charges Poisson commute with one another is known from \cite{Vicedo:2019dej}, in which Vicedo found that the 4d CS Poisson algebra (in the Lax gauge) is the Maillet bracket. 
    For our own ease, we will call $\mathcal{A}$ the Lax connection from here on in. 

    \subsubsection{\poletype{}{d}{s}} 

    Let us now turn our focus to the \poletype{}{d}{s} and the solution to the bulk equations of motion $\omega \wedge \cdel \mathcal{A} = 0$. 
    At a zero $\zeta \in Z$ we define \textit{\poletype{t}{d}{}} of type $(m_\zeta^+,m_\zeta^-)$ via the conditions: 
    \begin{equation}
        (z-\zeta)^{m^+_\zeta} A_{+} \ \ \text{ and } \ \ (z-\zeta)^{m^{-}_{\zeta}}A_{-} \ \ \text{ are regular, } \label{eq: Z-defect definition}
    \end{equation}
    where the pair of integers $(m_{\zeta}^{+},m_{\zeta}^{-})$ satisfy the inequality $m_{\zeta}^{+} + m_{\zeta}^{-} \leqslant m_\zeta$. 
    To preserve the defect type, we require that gauge transformations leave these conditions unchanged. 
    We do this by considering only those gauge transformations that are regular at the zeros of $\omega$. 

    As given in \cite{Delduc:2019whp} and \cite{Benini:2020skc}, solutions to $\omega \wedge \widebar{\partial} \lax{A} = 0$ when \poletype{}{d}{s} are inserted have a partial fraction expansion of the form:
	\begin{equation}
		\lax{A}_{\pm} = \lax{A}^{c}_{\pm}(x^{+},x^-) + \left( \sum_{\zeta \in Z \setminus \{\infty\}} \sum_{l=0}^{m^{\pm}_{\zeta}-1} \frac{\lax{A}^{\zeta , l}_{\pm}(x^+, x^-)}{(z-\zeta)^{l+1}}  \right) + \sum_{l=0}^{m^{\pm}_{\infty}-1} \lax{A}_{\pm}^{\infty , l}(x^+,x^-) z^{l+1}  \fp \label{Generic Lax}
	\end{equation}
	where $\lax{A}^{c}_{\pm},\lax{A}_{\pm}^{\infty , l},\lax{A}^{\zeta , l}_{\pm} : \Sigma \rightarrow \mathfrak{g}$ and in which the pair $(m_{\zeta}^+,m_{\zeta}^-)$ determine the allowed singular behaviour of $\mathcal{A}$'s components. 
    We shall see later on that the coefficients of this expression can be fixed in terms of a $\sigma$-model's fields using the boundary conditions introduced in the next subsubsection.  

    Rather than simply stating \eqref{Generic Lax}, we feel obliged to explain why $\lax{A}$ is of this form. To do this, we consider the behaviour of the equation: 
    \begin{equation}
		\omega \wedge \widebar{\partial} \lax{A}_{\pm} = 0 \ca \label{2.2}
	\end{equation}
    in two scenarios: $i)$ $z = \zeta$ and $ii)$ $z \neq \zeta$, where $\zeta \in Z$. 
    In scenario $i)$ the equality holds only if $\widebar{\partial} \lax{A}_{\pm} = 0$ which implies that $\lax{A}_{\pm}$ is not a function of $\bar{z}$.
    However, in scenario $ii)$ the equality in \eqref{2.2} can hold even if $\widebar{\partial} \lax{A}_{\pm} \neq 0$. 
    This is for the following reason. 
    Suppose that the zero $\zeta$ of $\omega$ is of order $m_\zeta$ and that $\mathcal{A}_{\pm}$ has a pole located at $\zeta$ of degree $m_{\zeta}^{\pm}$. 
    The $\bar{z}$-derivative of this pole will produce a delta function as $\cdel (z-\zeta)^{-1} = 2 \pi i \delta^2(z-\zeta)$, and we thus find that the resulting contribution to \eqref{2.2} is: 
    \begin{equation}
        2 \pi i \, \omega \sum_{\zeta \in Z} \delta^2(z-\zeta)\sum_{l=0}^{m_{\zeta}^{\pm}-1} (l+1)!  \frac{\mathcal{A}_{\pm}^{\zeta,l}(x^{+},x^-)}{(z-\zeta)^l}  \fp
    \end{equation} 
    which vanishes (upon integration) because $\omega(\zeta) = 0$ and $m_{\zeta}^{\pm} \leqslant m_{\zeta}$. 
    This would not be true if $m_{\zeta}^{\pm}$ were greater than $m_{\zeta}$, meaning that allowed degrees of $\mathcal{A}$'s poles are determined by the orders of $\omega$'s zeros. 
    It thus follows that $\mathcal{A}_{\pm}$ is meromorphic in $z$ with poles at the zeros of $\omega$. 
    Since meromorphic functions on $\mathbb{C} P^1$ are ratios of two polynomials in $z$, one finds that the generic solution is the partial fraction expansion \eqref{Generic Lax}. 
    The polynomial terms of \eqref{Generic Lax} follow from the assumption that $\omega$ has a zero at $z= \infty$, these terms are clearly poles by the inversion $z \rightarrow 1/z$.
	
	When $\varphi(z)$ is non-vanishing at infinity, \eqref{Generic Lax} reduces to:
	\begin{equation}
		\lax{A}_{\pm} = \lax{A}_{\pm}^{c}(x^+,x^-) + \sum_{\zeta \in Z} \sum_{l=0}^{m^{\pm}_{\zeta}-1} \frac{\lax{A}^{\zeta , l}_{\pm}(x^+,x^-)}{(z-\zeta)^{l+1}}  \fp \label{Lax solution}
	\end{equation}
    This includes our constructions because $\omega$ always has a pole at infinity. 

	\subsubsection{\zerotype{T}{D}{s}: Boundary Conditions on $A$}

	\label{boundary conditions in cwy} 
    
    In this section, we introduce three classes of ``P-defects'' first given in \cite{Costello:2019tri}, that are defined by certain boundary conditions at the poles of $\omega$. 
    They can be found by further simplifying equation \eqref{beom 1} under the requirement that each of the integrands in the sum over $q \in P$ vanishes separately. 
    When taken together with the fact that $\omega$ having at most double poles truncates the sum over $l$ to $l \leq 1$, this restriction produces the \textit{defect equations of motion}:
	\begin{equation}
		\left( k_{q}^{(0)} + k_{q}^{(1)} \partial_{z} \right) \tr{A(\textbf{\textit{q}})}{ \delta A(\textbf{\textit{q}}) } = 0  \ca \label{beom 2}
	\end{equation}
	in which $k_{q}^{(1)}$ = 0 for a simple pole, and where we have introduced the notation\footnote{We leave it as implicit that the $z$-derivative acts on the expression $\tr{A}{ \delta A }$ before its evaluated at $\textbf{\textit{q}}$.} $f(\textbf{\textit{q}}) := f(x^{\pm}, q, \bar{q})$. 
    Notice that a solution of \eqref{beom 2}, and thus each \zerotype{}{d}{} is associated with a single pole of $\omega$. 

    Importantly, the boundary conditions found by solving equation \eqref{beom 2} must be unchanged by the gauge transformations of $A$:
	\begin{equation}
		A \rightarrow {}^u A := u (d + A) u^{-1} \fp \label{Eq: A gauge transform.}
	\end{equation}
    This requirement produces a set of constraints on the group element $u : \man \rightarrow G$ that will be useful when discussing the gauge invariance of the standard action \eqref{cwy action}. 
    We shall call defect-preserving transformations \textit{admissible gauge transformation}. 

    As discussed earlier, the 4d CS action is independent of any metric on $\Sigma$, but the two-dimensional action to which it is equivalent is either Euclidean or Lorentzian.
    The signature of the action is determined by the nature of the boundary conditions we impose.
    For simplicity, we discuss the Lorentzian case, in which we take the real coordinates $x^\pm$ on $\Sigma$ that will eventually become light-cone coordinates of a Lorentzian $\sigma$-model.
    The Euclidean case is easily achieved by substituting complex coordinates $(u,\bar u)$ for $(x^+,x^-)$.

    The three boundary conditions that we consider are given in \ref{table:1} together with the required constraint upon $u$. 
    The first two of these, which we will call chiral and anti-chiral\footnote{This nomenclature is chosen because one finds a chiral/anti-chiral Kac-Moody current on the associated defect.}, are imposed at first-order poles in $\omega$, while the third, simply called Dirichlet, is imposed at a second-order pole. 
    Notice that the potentially singular term in equation \eqref{eq: singular term} is finite for Dirichlet conditions. 
    Finally, this list is not exhaustive; others are discussed elsewhere, see \textit{e.g.} \cite{Costello:2017dso,Delduc:2019whp}.

    \begin{table}[h!]
    \centering 
    \def\arraystretch{1.5}
    \begin{tabular}{ c | c | c | c } 
        Defect Type & Degree of $\omega$'s Pole & Boundary Condition & Constraint on $u$ \\
        \hline
        Chiral & 1 & $A_{-}(\textbf{\textit{q}}) = 0$ & $\partial_{-} u(\textbf{\textit{q}}) = 0$ \\ 
        \hline
        Anti-chiral & 1 & $A_{+}(\textbf{\textit{q}}) = 0$ & $\partial_{+} u(\textbf{\textit{q}}) = 0$ \\
        \hline
        Dirichlet & 2 & $A_{\pm}(\textbf{\textit{q}}) = 0$ & $\partial_{\pm} u(\textbf{\textit{q}}) = 0$ \\
        \end{tabular} 
        \caption{The defect conditions used in standard 4d CS.}
        \label{table:1}
    \end{table}

	\subsection{Gauge Invariance}

	\label{section:cwy-gauge-invariance}

	We have already discussed the unusual gauge invariance of the four-dimensional action; we are now in a position to discuss invariance of the action under admissible gauge transformations $A \rightarrow {}^u A$ that preserve the boundary conditions at poles of $\omega$. 
    Under such gauge transformations, we find the action \eqref{cwy action} transforms as:
	\begin{equation}
		S_{\text{4dCS}}(A) \rightarrow S_{\text{4dCS}}({}^{u} A) = S_{\text{4dCS}}(A) + \frac{i}{8 \pi^2} \int_{M} \omega \wedge \cdel \tr{u^{-1} d u}{A}  + S_{\text{WZ}}(u) \fp \label{cwy gauge transformed}
	\end{equation}
    where: 
    \begin{equation}
        S_{\text{4dWZ}} (u) = \frac{i}{48 \pi^2} \int_{M} \omega \wedge \tr{u^{-1} d u}{[u^{-1} d u,u^{-1} du]} \fp \label{wess-zumino term}
    \end{equation}
    When deriving the above equations, we have sent a total derivative of $d_{\Sigma}$ to zero by requiring $A$ to die off to zero, and $u$ to the identity, at infinity in $\Sigma$. 
    In the following, we denote the second term on the right-hand side by $\delta_u S_{\partial}$ and require that both it and the third term $S_{\text{4dWZ}}$ vanish separately; we consider $\delta_u S_{\partial}$ first.

    Using \eqref{integral over poles} with $\xi = \tr{u^{-1} du}{A}$, and noting that each pole is at most second order, we find:
	\begin{equation}
		\delta_u S_{\partial} = \frac{1}{4 \pi}  \sum_{q \in P} \delta_u S_{q}
		\;,\;\;\;\;
		\delta_u S_{q} = \int_{\Sigma_{q}} \left( k_{q}^{(0)} + k_{q}^{(1)} \partial_{z} \right)\tr{u^{-1}_q d u_q}{A} \label{gauge trans. term by term} \fp
	\end{equation}
    where $u_q = u \pp$.
    For the action to be gauge invariant, the above expression must vanish.
    This is indeed the case for the three boundary conditions defined in table \ref{table:1}, and happens precisely because of the constraints imposed upon $u_q$.

	\paragraph{Chiral boundary conditions:}

	We take $\omega$ to have a simple pole at $z=q$ (where $k_{q}^{(1)} = 0$) and impose the chiral boundary condition $A_{-} \pp = 0$, thus reducing equation \eqref{gauge trans. term by term} to:
	\begin{equation}
		\delta_u S_{q} = - k^{(0)}_{q} \int_{\Sigma_{q}} d^2x \tr{u^{-1}_q \partial_{-} u_q }{ A_{+} } = 0\ca
    \end{equation}
	where $d^2x = d  x^{+} \wedge d x^{-}$ and with the final equality holding because of the constraint $\partial_{-} u_q = 0$. 
    Hence, any contribution due to a first-order pole in the second term of equation \eqref{cwy gauge transformed} can be made to vanish when imposing chiral boundary conditions.

	\paragraph{Anti-chiral boundary conditions:}

	We take $\omega$ to have a simple pole at $z=q$ and impose the anti-chiral boundary condition $A_{+} \pp = 0$. 
    After imposing this, the term $\delta_u S_{q}$ in equation \eqref{gauge trans. term by term} vanishes upon imposing the constraint $\partial_{+} u_q = 0$. Hence, any contribution due to a first-order pole in the second term of equation \eqref{cwy gauge transformed} can be made to vanish upon imposing anti-chiral boundary conditions.

	\paragraph{Dirichlet boundary conditions:}

	Finally, we take $\omega$ to have a double pole at $z=q$, at which we impose the Dirichlet boundary conditions, hence \eqref{gauge trans. term by term} is:
	\begin{equation}
		\delta_u S_{q} = \int_{\Sigma_{q}} ( k^{(0)}_{q} + k^{(1)}_{q} \partial_{z} ) \tr{u^{-1}_q d u_q }{ A}= 0 \label{dirichlet gauge inv.} \fp
	\end{equation}
	The condition $A_{\pm} \pp = 0$ means the first term in equation \eqref{dirichlet gauge inv.} vanishes. This leaves us with:
	\begin{equation}
		\delta_u S_{q} = - k^{(1)}_{q} \int_{\Sigma_{q}}  d^2 x \, \partial_{z} \epsilon^{jk} \tr{ u^{-1}_q \partial_{j} u_q }{A_{k}} \ca
	\end{equation}
	for $j,k = \pm$ and where $\epsilon^{+-}=1$. 
    Upon imposing the constraints $\partial_{\pm} u_q = 0$ and $A_{\pm} \pp = 0$ together, we find that this term also vanishes. 
    Hence, any contribution due to a second-order pole vanishes when we impose a Dirichlet boundary condition.

	
	\paragraph{The Wess-Zumino Term:}
	
	The final step in proving gauge invariance is to show the Wess-Zumino term \eqref{wess-zumino term} vanishes. 
    To do this, we first take the exterior derivative of the Wess-Zumino three-form and find that it is closed:
	\begin{equation}
		d \tr{ u^{-1} d u}{[u^{-1} d u,u^{-1} du]} = - \tr{(u^{-1} d u)^2}{[u^{-1} d u,u^{-1} du]} = 0 \ca \label{WZ closure}
	\end{equation}
	because of the cyclic symmetry of the trace. 
	Since the three-form is closed, it is natural to ask whether it is exact. We can answer this by calculating the third de Rham cohomology of $\Sigma\times \mathbb{CP}^1$, $H^{3}_{\text{dR}}(\Sigma \times \mathbb{C} P^{1})$.
	As we will fix $\Sigma = \mathbb{R}^2$ throughout the following, we restrict ourselves to this case only. 
    We can therefore use the K{\"u}nneth theorem and the cohomologies of $\mathbb{R}^2$ and $\mathbb{CP}^{1}$ (see appendix \ref{appendix:DeRham}) to find that $H^{3}_{\text{dR}}(\mathbb{R}^2 \times \mathbb{C} P^{1})=0$, which implies the Wess-Zumino three-form is exact on $\mathbb{R}^2 \times \mathbb{C} P^{1}$: 
    \begin{equation}
        \tr{ u^{-1} d u}{[u^{-1} d u,u^{-1} du]} = d E(u) \fp
    \end{equation}

    To find an expression for $E(u)$ we introduce the coordinates $\{\zeta^a\}$ on $G$ and define the anti-symmetric tensor $\lambda_{ab}$ by: 
    \begin{equation}
        \tr{ u^{-1} \partial_a u}{[u^{-1} \partial_b u,u^{-1} \partial_c u]} = \partial_{a} \lambda_{bc} + \partial_{b} \lambda_{ca} + \partial_{c} \lambda_{ab} \ca
    \end{equation}
    where $\partial_a = \partial / \partial \zeta^a$. 
    The Wess-Zumino three-form is thus: 
    \begin{equation}
        \tr{ u^{-1} d u}{[u^{-1} d u,u^{-1} du]} = 3 \partial_c \lambda_{ab} d \zeta^c \wedge d \zeta^a \wedge d \zeta^b = 3 d( \lambda_{ab} d \zeta^a \wedge d \zeta^b) \fp
    \end{equation}
    If we plug the above equation into the Wess-Zumino term $S_{\text{4dWZ}}(u)$ we find an expression of the form \eqref{eq: boundary terms}, with $\xi = \lambda_{ab} d \zeta^a \wedge d \zeta^b$, which we can then collapse to a sum over poles using equation \eqref{integral over poles}: 
    \begin{align}
        S_{\text{4dWZ}}(u) = \frac{1}{2 \pi} \sum_{q \in P} \sum_{l=0}^{n_q-1} \frac{k_q^{(l)}}{l!} \int_{\Sigma_q} \partial_z^l (\lambda_{ab} d \zeta^a \wedge d \zeta^b) \fp
    \end{align}
    Whether $q$ is a simple or double pole of $\omega$, the above expression vanishes for all three boundary conditions given in table \ref{table:1}. 
    This is because $\partial_{i}u = \partial_i \zeta^a \partial_a u = 0$ for either $i=+$ or $i=-$ or both. 
	
\section{Integrable Sigma Models on \zerotype{T}{D}{s}}

	\label{section: Integrable Sigma Models} 

    In this section, we review the ``DLMV construction'' of \cite{Delduc:2019whp}, in which the authors found a unified action for a class of integrable $\sigma$-models. 
    The starting point of this derivation is to simultaneously change variables and partially gauge fix by going from $A$ to a group element $\hat \sigma$ and the Lax connection $\mathcal A$. 
    This is done via the equations\footnote{These are occasionally described as a ``formal'' gauge transformation of the Lax connection $\mathcal{A}$ to $A$. This language is chosen to distinguish it from an admissible gauge transformation, as $\hat{\sigma}$ will not, in general, preserve boundary conditions -- if it did, there would be fewer physical degrees of freedom. It is also for this reason that we do not work in the Lax gauge.}: 
    \begin{equation}
        \widebar{A} = \hat \sigma \cdel \hat \sigma^{-1} \ca \qquad A_\Sigma = \hat \sigma d_\Sigma \hat \sigma^{-1} + \hat \sigma \mathcal A \hat \sigma^{-1} \label{eq: change of variables}
    \end{equation}
    As was observed by Costello and Yamazaki, there is, in fact, a whole class of group elements $\{\hat{\sigma}\}$ that are each defined by the first of these two equations. 
    The next step is to impose the ``archipelago'' gauge by selecting a canonical element $\tilde{g}$ from the set $\{ \hat{\sigma} \}$ that satisfies a set of conditions, called ``archipelago'' conditions. 
    These will be introduced below. 
    The final step in this derivation is to substitute into the 4d CS action the expressions given in equation \eqref{eq: change of variables}, with $\hat \sigma = \tilde g$. 
    Using equation \eqref{integral over poles} and the archipelago condition, we find the ``unified $\sigma$-model action''.

	\subsection{The Class of $\sigma$-Model Fields}

	\label{vicedo's construction}
	
	In \cite{Costello:2019tri}, Costello and Yamazaki proved the existence of a class of group elements $\{\hat{\sigma}\}$ such that:
	\begin{equation}
		A_{\bar{z}} (x^{\pm},z, \bar{z}) = \hat{\sigma} \partial_{\bar{z}} \hat{\sigma}^{-1} \label{ghat def} \fp
	\end{equation}
	where $\hat{\sigma} : \man \rightarrow G$. 
    This is analogous to the construction of the Wess-Zumino-Witten model in \cite{Elitzur:1989nr} from 3d CS, but here $\hat{\sigma}$ is not found explicitly as a path-ordered exponential. 
    The equation \eqref{ghat def} has a right acting symmetry transformation, called the \textit{right-redundancy}:
	\begin{equation}
		\hat{\sigma} \rightarrow \hat{\sigma}^{\prime} = \hat{\sigma} k_g \ca \label{right gauge symmetry}
	\end{equation} 
    where $\partial_{\bar z} k_g = 0$, meaning that $\hat \sigma$ and $\hat \sigma'$ give the same $A_{\bar z}$. 
    Since $\mathbb{C} P^{1}$is compact, any holomorphic function is constant and thus $k_g$ is a function of $x^{\pm}$ only. 

	We fix this right-redundancy by choosing a canonical group element $\hat g$ in the class $\{\hat{\sigma}\}$ which is defined to satisfy $\hat{g}_{\infty} = 1$.
	One can always find $\hat{g}$ from any element $\hat{\sigma} \in \{\hat{\sigma}\}$ via a right redundancy transformation \eqref{right gauge symmetry} by $k_g  = \hat{\sigma}^{-1}_{\infty}$:
	\begin{equation}
		\hat{g}(x^{\pm}, z, \bar{z}) = \hat{\sigma} (x^\pm,z,\bar z)\cdot \hat{\sigma}
		(x^\pm,\infty,\infty)^{-1} \ca \label{gpi def}
	\end{equation}
	where clearly $\hat{g}_{\infty} = 1$. 
	Since we can invariably select $\hat{g}$ from $\{\hat{\sigma}\}$ we will do so throughout the following. 
	
	\subsubsection{The Archipelago Conditions}
	
	\label{section: Archipelago Conditions}
	
	Here, we correct a minor error in \cite{Delduc:2019whp} and prove that there exists a gauge, called the ``archipelago'' gauge, in which a group element $\tilde{g} \in \{\hat{g}\}$ satisfies the ``archipelago'' conditions of \cite{Delduc:2019whp}. 
    This is expressed in terms of $\tilde g$, but since $A_{\bar z}$ is determined by $\tilde g$, this is a partial gauge choice on $A$. 	
    
    These archipelago conditions are the following.
	For each pole $q$ in $P$, let $U_q$ be a disc of radius $R_q$. 
    If $q$ is in $P_{\text{fin}}$ then $U_{q}$ is the region of $\mathbb{C} P^1$ that satisfies $|z-q| < R_{q}$; while for the pole at infinity, $U_\infty$ is the region  $|1/z| < R_{\infty}$. 
    We require that the radii $R_{q}$ be chosen to ensure that these discs are disjoint. 
    Using the discs, we define the \textit{archipelago conditions} by:
	\begin{itemize}
		\item[($i$)] $\tilde{g} = 1$ outside the disjoint union $\Sigma \times \sqcup_{q \in P} U_{q}$;
		\item[($ii$)] Within each $\Sigma \times U_{q}$ we require that $\tilde{g}$ depends only upon $x^\pm$ and the radial coordinate $r_{q}$ of the disc $U_{q}$. We choose the notation $\tilde{g}_{q}$ to indicate that $\tilde{g}$ is in the disc $U_{q}$, this condition means that $\tilde{g}_{q}$ is rotationally invariant;
		\item[($iii$)] For every $q \in P$ there is an open disc $V_{q} \subset U_{q}$ of radius $S_q < R_q$ which is centred on $q$. In the disc $V_q$ $\tilde{g}_{q}$ depends upon $x^{+}$ and $x^{-}$ only. We denote $\tilde{g}_{q}$ in this region by $g_{q} = \tilde{g}|_{\Sigma \times V_{q}}$.
	\end{itemize}
	The first of these condition sets $A_{\bar z}=0$ outside $\Sigma \times \sqcup_{q \in P} U_{q}$, the second ensures that $A_{\bar{z}}$ is rotationally invariant in $U_{q}$, and the third sets  $A_{\bar z}=0$ inside $\Sigma \times \sqcup_{q \in P} V_{q}$.

	\begin{figure}[h]
		\centering
		\begin{tikzpicture}
			\node at (0,0) {\pgfuseimage{many-poles}};
			\node at (-3.25,3.25) {$\mathbb{C} P^{1}:$};
			\node at (0.1,-3.3) {$\tilde{g}=g_{0}$};
			\node at (0.1,3.25) {$\tilde{g}=1$};
			\node at (3.25,1.4) {$\tilde{g}=g_{1}$};
			\node at (0.8,1.5) {$\tilde{g}=g_{2}$};
			\node at (-1.2,1.6) {$\tilde{g}=g_{3}$};
			\node at (-0.9,-1.6) {$\tilde{g}=g_{4}$};
			\node at (1.8,-1.1) {$\tilde{g}=g_{5}$};
		\end{tikzpicture}
		\caption{An illustration of the archipelago conditions for an $\omega$ with seven poles and five zeros. The diamonds represent the poles of $\omega$ with the enclosing circles illustrating the discs $U_q$. Each $g_i = \tilde{g} |_{q_i}$ denotes the value of $\tilde{g}$ at the associated pole of $\omega$. The five black triangles represent the zeros of $\omega$ at which $A$ can have poles.}
	\end{figure}
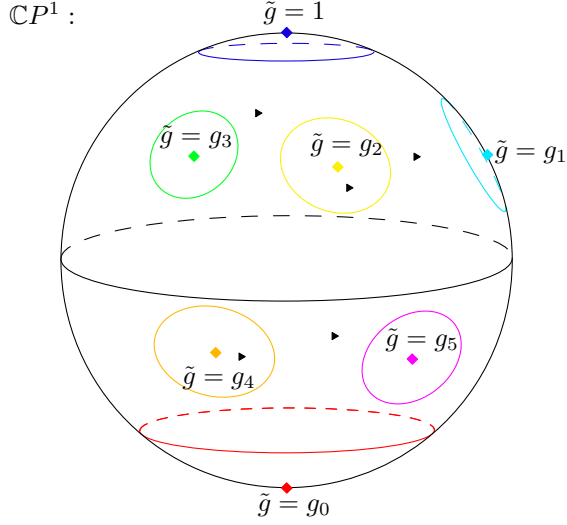
	
	We now prove the existence of the archipelago gauge in two steps. 
    In the first, we show how to construct a $\tilde{g}$ satisfying the above conditions from any $\hat{g}$; and in the second show that there exists a admissible gauge transformation $u = \tilde{g} \hat{g}^{-1}$, which puts $A$ into the archipelago gauge.  
	
    In \cite{Delduc:2019whp}, the authors proposed a construction of $\tilde{g}$ which satisfies the archipelago conditions, but their construction is not quite right because it involves expressing $\tilde{g}$ as an exponential of a Lie algebra element. 
    Although $\tilde{g}$ is in the identity component of $G$, by the first archipelago condition, it is not the case that $\tilde{g}$ can be constructed as an exponential everywhere in this component.
    For example, if we take $G = SL(2,\mathbb{C})$ then the group element:
	\begin{equation}
		\begin{pmatrix}
			-1 & 1 \\
			0 & -1
		\end{pmatrix} \ca
	\end{equation}
	is in the identity component of $SL(2,\mathbb{C})$ but cannot be written as an exponential of an element of the Lie algebra $\mathfrak{sl}_2 (\mathbb{C})$. 
    It is for this reason that the treatment presented below is slightly different to that presented in \cite{Delduc:2019whp}.

	This minor issue is easily solved by the following argument. Let $\hat g$ be the original group element and $\tilde g$ the group element in archipelago gauge.
    At each pole $q \in P$ we choose $g_q = \tilde g \pp$ to be equal to $\hat{g}$:
    \begin{align}
	    g_q = \hat g \pp \label{eq:3.6}
	    \;.
	\end{align}
    Since $g_{\infty} = \hat g|_{\infty}=1$ and $\tilde g$ varies smoothly over $M$, $\tilde g$ must be in the identity component of $G$ everywhere on $M$. This means, for each disc $U_q$, we can choose a smooth path in $G$ between $1$ and $g_q$, $\gamma_{q} (t,x^{\pm}) : [0,1] \times \sigpi \rightarrow G$ where $\gamma_q (0,x^{\pm}) = 1$ and $\gamma_q (1,x^{\pm}) = g_q$. 

    To construct the smooth path in $U_q$ we use a bump function, which by definition is smooth with compact support in the domain.
    For each $U_q$ let $W_q \supset U_q$ be a disc of radius $R_q + \epsilon$ (where $\epsilon > 0$) which, by an abuse of notation, has the radial coordinate $r_q$ and centre $U_q$ on $r_q = 0$.
    We define for each $W_q$, a bump function $f_q : W_q \rightarrow \mathbb{R}$ with the following properties:
    \begin{equation}
        f_q(r_q) = \begin{cases}
            f_q(r_q) = 1 \text{ if } r_q < S_q \ca \\
            f_q(r_q) = 0  \text{ if } r_q \geqslant R_q \ca \\
            f_q(r_q) \text{ interpolates smoothly between }  1 \text{ and } 0 \text{ if } S_q \leqslant r_q < R_q \fp 
        \end{cases}
    \end{equation}
    Hence, in the disc $W_q$ we can define $\tilde{g}$ by:
    \begin{equation}
        \tilde{g}(r_q,x^{\pm}) = \gamma_q (f_q(|z-q|),x^\pm) \fp
    \end{equation}
    This satisfies conditions $(ii)$ and $(iii)$ of the archipelago conditions. 
    In $U_{\infty}$ we have $\tilde{g} (r_{\infty} = 0,x^{\pm}) = \tilde{g} (r_{\infty}=R_{\infty},x^\pm) = 1$ and so choose a path where $\tilde{g} (r_{\infty},x^{\pm})=1$ for all values of $r_{\infty}$.

    Globally, we define $\tilde{g}$ by:
    \begin{equation}
        \tilde g(z,\bar z,x^{\pm})
	    = 
	    \begin{cases}
	    1 & z \in \remsph \setminus \sqcup_{q \in P_{\text{fin}}} W_{q} \ca \\
	   \gamma_q (f_q(|z-q|),x^\pm) & 
	    z \in W_q \ca
	    \end{cases} 
	    \label{eq:gtildef}
    \end{equation}
    which satisfies all three archipelago conditions and is smooth by construction. 

    To enter the archipelago gauge, we must be able to perform an admissible gauge transformation from $\hat{g}$ to $\tilde{g}$. 
    The required transformation is generated by $u = \tilde{g} \hat{g}^{-1}$. 
    It preserves the boundary conditions given in table \ref{table:1} because equations \eqref{eq:gtildef} and \eqref{eq:3.6} imply that $u = 1$ at all poles $q \in P$, and therefore that $\partial_{\pm} u = 0$. 
    Thus, we will use the following equations for $A$: 
    \begin{equation}
        \widebar{A} = \tilde g \cdel \tilde g^{-1} \ca \quad A_\Sigma = \tilde g d_\Sigma \tilde g^{-1} + \tilde g \mathcal A \tilde g^{-1} \label{eq: A in the archipelago gauge}
    \end{equation}
    for the remainder of this section.

    \subsubsection{Symmetry Transformations of the Lax Connection \label{section: Symmetries of the Lax connection}} 

    Let's discuss the two classes of transformations induced upon the pair $(\tilde g ,\mathcal A)$ by $i)$ the right-redundancy and $ii)$ residual gauge transformations of $A$, \textit{i.e.} those transformations that leave us in the archipelago gauge. 

    In case $i)$, we know already that $\tilde{g} \rightarrow \tilde{g} k_g$, meaning we need only find the transformation law for $\mathcal A$. 
    To do this, we observe that every element of the class $\{ \hat \sigma \}$ must produce the same field configuration for $A$. 
    This is because $A_{\pm}$ is determined by $A_{\bar z}$ when we solve the equations of motion $\omega F_{\bar z \pm} = 0$, and each element of $\{ \hat{\sigma} \}$ gives the same $A_{\bar{z}}$. 
    Therefore, $A$ is invariant under the right-redundancy and we find: 
    \begin{align}
		\lax{A} =& \,  \tilde{g}^{-1} d \tilde{g} + \tilde{g}^{-1} A \tilde{g} \rightarrow  (\tilde{g} k_g)^{-1} ( d + A) (\tilde{g} k_g) \nonumber \\
		=& \, {k_g}^{-1}(\tilde{g}^{-1} d \tilde{g}+\tilde{g}^{-1} A \tilde{g} ) k_g + {k_g}^{-1} d k_g \nonumber \\
		=& \, {k_g}^{-1} \lax{A} k_g + {k_g}^{-1} d k_g \fp
	\end{align} 
    All together, the full right-redundancy transformation is: 
    \begin{equation}
		\tilde{g} \rightarrow \tilde{g} k_g, \quad \lax{A} \rightarrow {}^{k_g} \lax{A} := {k_g}^{-1} \lax{A} k_g + {k_g}^{-1} d k_g \label{rrL} \fp
	\end{equation} 
    This implies that each field configuration of $A$ is associated with a gauge equivalence class of Lax connections, and that there is no preferred Lax connection. 
    
    To derive case $ii)$, we need a transformation of $\tilde{g}$ under $u$ that reproduces $A \rightarrow u (d+A) u^{-1}$ while also leaving our fixing of the right-redundancy unchanged. 
    This latter condition is required because the right-redundancy parametrises our freedom to choose elements from the class $\{ \hat \sigma \}$ and must therefore be unaffected by gauge transformations. 
    In essence, we can fix the right-redundancy in the same way for both $A_{\bar z} = \tilde{g} \cdel \tilde{g}^{-1}$ and ${}^u A_{\bar{z}} = {}^u \tilde{g} \cdel ( {}^u \tilde{g})^{-1}$. 
    These two requirements lead us to:
    \begin{equation}
        \tilde{g} \rightarrow {}^u \tilde g =u \tilde{g} u_{\infty}^{-1} \ca \qquad \lax{A} \rightarrow {}^u \lax{A} = u_{\infty} \dsig u_{\infty}^{-1} + u_{\infty} \lax{A} u_{\infty}^{-1} \ca
    \end{equation}
    which is the symmetry of the $\sigma$-models discussed in the next two subsections. 

    \subsection{The Unified $\sigma$-Model \label{section: USM}}
	
	We now reduce the 4d CS action down to two dimensions.
    This is done in three steps. 
    In the first, we substitute in:
    \begin{equation}
		A = \tilde{g} d \tilde{g}^{-1} + \tilde{g} \lax{A} \tilde{g}^{-1} \ca \label{eq: archipelago imposed on A}
	\end{equation} 
    which is of the same form as a gauge transformation, allowing us to use equation \eqref{cwy gauge transformed} and find: 
    \begin{gather}
		S_{\text{4dCS}}(A) = \frac{i}{8 \pi^2} \int_{M} \omega \wedge \tr{\lax{A} }{\cdel \lax{A}} + \frac{i}{8 \pi^2} \int_{\man} \omega \wedge \cdel \tr{ \tilde{g}^{-1} d \tilde{g}}{ \lax{A} } + S_{\text{4dWZ}}(\tilde{g}) \label{reduced action3} \fp
	\end{gather}
    The first term is just $S_{\text{4dCS}}(\mathcal{A})$ since $\mathcal{A}_{\bar z} = 0$, by definition. 
    In the second step, we impose the equation of motion $\omega \wedge \cdel \mathcal{A} = 0$ and take $\mathcal A$ to be of the form \eqref{Lax solution}. 
    Doing this causes the first term in \eqref{reduced action3} to vanish because the action of $\cdel$ picks up delta functions at the locations $\zeta \in Z$, which, when integrated, impose $\omega(\zeta) = 0$. 
    This zero dominates any poles in $\mathcal A \wedge \cdel \mathcal A$ due to the inequality $m_\zeta^+ + m_\zeta^- \leq m_\zeta$. 
    
    In the final step, we use the archipelago conditions to integrate out any dependence on the angular coordinates of the discs $U_q$ introduced in the previous section. 
    Starting with the second term on the right-hand side of \eqref{reduced action3}, we use equation \eqref{integral over poles} to find:
    \begin{equation}
		\frac{i}{8 \pi^2} \int_{M} \omega \wedge \cdel \tr{ \tilde{g}^{-1} d \tilde{g}}{ \lax{A} } = \frac{1}{4 \pi} \sum_{q \in P_{\text{fin}}} \int_{\Sigma_{q}} \tr{ g_{q}^{-1} d g_{q}}{ \text{Res}_{q} (\varphi(z) \lax{A}) } \ca
	\end{equation}
	where archipelago condition $iii)$ allows us to remove $g_{q}^{-1} d g_{q}$ from the residue. 
    This sum is restricted to be over $P_{\text{fin}}$ as $\tilde g_{\infty}=1$ implies $\tilde g_{\infty}^{-1} d \tilde g_\infty = 0$. 
    
    The reduction of the four-dimensional Wess-Zumino term has to be done in several steps.
    In the first, we use archipelago condition $i)$ to decompose it into a sum of integrals:
    \begin{align}
        S_{\text{4dWZ}}(\tilde{g}) = \sum_{q \in P_{\text{fin}}} S_{\text{WZ}} (\tilde g_q) \ca \quad \text{where} \quad S_{\text{WZ}} (\tilde g_q) =  \frac{i}{48 \pi^2} \int_{\Sigma \times U_q} \! \! \! \! \! \! \! \! \! \omega \wedge \tr{\tilde g_q^{-1} d \tilde g_q}{[\tilde g_q^{-1} d \tilde g_q,\tilde g_q^{-1} d \tilde g_q]} \ca
    \end{align}
    and with the sum again restricted to be over $P_{\text{fin}}$ for the reason we've just given. 
    To evaluate each of these integrals, we introduce a local set of polar coordinates $(r, \theta)$ on $U_q$ defined by $z-q=r \exp( i \theta)$, and use archipelago condition $ii)$, which implies $\tilde{g}^{-1} \partial_{\theta} \tilde{g} = 0 $. 
    Upon doing this, we find: 
    \begin{align}
        S_{\text{WZ}} (\tilde g_q) = - \frac{ k_{0}^{(-1)} }{48 \pi^2}\int_{\Sigma \times U_{q}} \! \! \! \! \! \! \! r e^{i \theta} d \theta \wedge & \tr{\tilde g_q^{-1} \tilde d \tilde g_q}{[\tilde g_q^{-1} \tilde d \tilde g_q,\tilde g_q^{-1} \tilde d \tilde g_q]} \label{eq: intermediate WZ} \\ 
    &- \frac{1}{4 8 \pi^2} \! \! \sum_{q' \in P_{\text{fin}}} \sum_{l=0}^{n_{q'}-1}  \frac{k_{q'}^{(l)}}{l!} \int_{M_q} \frac{1}{r^{l+1}}I^{(l)}_{qq'}(r) \tr{\tilde g_q^{-1} \tilde d \tilde g_q}{[\tilde g_q^{-1} \tilde d \tilde g_q,\tilde g_q^{-1} \tilde d \tilde g_q^]} \nonumber
    \end{align} 
    where $\tilde d = dr \partial_r + dx^{+} \partial_{+} + dx^- \partial_{-}$, $M_q = \Sigma_q \times [0,R_q]$ and:
    \begin{equation}
        I^{(l)}_{qq'}(r) = \int_{0}^{2 \pi} d \theta \frac{e^{i \theta}}{ \left( e^{i \theta} + \tfrac{q-q'}{r} \right)^{l+1}} \ca
    \end{equation} 
    The first term on the right-hand side of equation \eqref{eq: intermediate WZ} goes to zero after integrating over $\theta$ while:
    \begin{equation}
        I^{(l)}_{q q'} (r) = \begin{cases}
            2 \pi \quad \text{if} \ \ q=q' \ \ \text{and} \ \ l=0 \, ,\\
            0 \ \ \ \ \ \text{otherwise} \fp
        \end{cases}
    \end{equation}
    For $q \neq q'$ and $l=0$, this is because $|q-q'|/r > 1$ as we have $|q-q'| > R_q$ and $r \in [0,R_q]$ by construction. 
    Thus: 
    \begin{equation}
        S_{\text{WZ}} (\tilde g_q) = -\frac{k_q^{(0)}}{24 \pi} \int_{M_q} \! \! \! \tr{\tilde g^{-1}_q d \tilde g_q}{[\tilde g^{-1}_q d \tilde g_q,\tilde g^{-1}_q d \tilde g_q]} \fp
    \end{equation}
    
    After putting all of this together we find equation \eqref{reduced action3} reduces to the \textit{unified} $\sigma$-\textit{model}: 
    \begin{equation}
         S_{\text{U} \Sigma \text{M}} (\tilde g,\mathcal A) := S_{\text{4dCS}} (A) = \sum_{q \in P_{\text{fin}}} \left( \frac{1}{4 \pi} \int_{\Sigma_{q}} \tr{ g_{q}^{-1} d g_{q}}{ \text{Res}_{q} (\varphi(z)\lax{A}) } + S_{\text{WZ}}(\tilde{g}_q) \right) \ca \label{unified sigma model action}
    \end{equation}  
    whose equations of motion are: 
    \begin{equation}
        d_\Sigma \mathcal A + \frac{1}{2} [\mathcal A, \mathcal A] = 0 \fp 
    \end{equation} 

    Since this action was found from the gauge invariant 4d CS action, it is invariant under the gauge transformations of $A$ that preserve the archipelago conditions, as well as the right-redundancy (which, as explained earlier, leaves $A$ unchanged). 
    This was shown explicitly in \cite{Delduc:2019whp}. 

\subsection{Examples of Models}

    Let's use the unified $\sigma$-model action to construct several different IMs. 
    Doing this will be particularly insightful, as it enables us to illustrate several important subtleties of these methods. 
    The most significant of these is that given an $\omega$, not every combination of defects produces an interesting $\sigma$-model. 
    The four cases we consider are: 
    \begin{enumerate}
        \item[I:] $\omega = dz$ with a Dirichlet P-defect,
        \item[II:] $\omega = \dfrac{k}{z} dz$ with two chiral P-defects,
        \item[III:] $\omega = \dfrac{k}{z} dz$ with a chiral and anti-chiral P-defect, 
        \item[IV:] $\omega =  \dfrac{z^2-\lambda^2}{(z+k)^2} dz$ with two Dirichlet P-defects and a pair of $(1,0)$ and $(0,1)$ Z-defects. 
    \end{enumerate}
    For simplicity's sake, we specialise to the case where $\Sigma = \mathbb{R}^2$, and work with the light-cone coordinates $x^{+}$, and $x^{-}$. The metric and volume forms will be $\eta^{+-} = 2, \eta^{++} = \eta^{--} = 0$ and $d^2 x = dx^{+} \wedge dx^{-}$.

    The general structure of the argument is as follows. 
    Suppose we have chosen $\omega$ and specified a defect configuration.  
    For each $q \in P$, we substitute the boundary conditions of the associated defect into $\mathcal A_i \pp = g_q^{-1} A \pp g_q +g^{-1}_q \partial_i g$ and find: 
    \begin{equation}
    	\lax{A}_{i} \pp = g^{-1}_{q} \partial_{i} g_{q} \label{bc equation} \ca
    \end{equation} 
    where $i=-$ if the defect is chiral, $+$ if it is anti-chiral or both if Dirichlet. 
    After doing this for all the poles, we find a set of simultaneous equations that define a linear algebra problem of the form $N X = Y$, where $X$ and $Y$ are vectors and $N$ is a matrix. 
    The elements of $X$ and $Y$ are, respectively, the coefficients of $\mathcal A$ given in \eqref{Lax solution} and the currents $g_q^{-1} \partial_{\pm} g_q$. 
    The exact structure of $N$ depends on the defect configuration under consideration and is best understood via examples. 
    If $N$ is invertible, then all coefficients of $\mathcal{A}$ can be determined in terms of $g_q$'s by $X = N^{-1} Y$ and we have a Lax connection for a particular $\sigma$-model. 
    The action of this model is found by substituting the Lax connection into the unified $\sigma$-model and evaluating the residues. 

\subsubsection{I: The Trivial Field Configuration} 

    \begin{table}[h!]
        \centering
        \def\arraystretch{1.5}
        \begin{tabular}{c|c|c}
             $\omega$ & \multicolumn{2}{|c}{P-defects}  \\ 
            \hline
             \multirow{2}{*}{$dz$} & Type & Position in $\mathbb{C}P^1 $ \\ 
             \cline{2-3}
             & Dirichlet & $\infty$
        \end{tabular} 
        \caption{The defect configuration in example I.}
        \label{table:2}
    \end{table}

    The defect configuration for this example is given in Table \ref{table:2}. 
    Since $\omega$ has no zeros, we do not introduce any Z-defects, and the solution \eqref{Lax solution} for $\mathcal A$ simplifies to: 
    \begin{equation}
        \lax{A} = \lax{A}_{+}^{c}(x^{+},x^{-}) dx^{+} + \lax{A}_{-}^{c}(x^{+},x^{-}) dx^{-} \fp
    \end{equation}
    Once substituted into equation \eqref{bc equation}, the boundary condition at infinity and $g_{\infty} = 1$ imply that: 
    \begin{equation}
        \mathcal{A}^{c}_{\pm} = 0 \ca
    \end{equation}
    One does not recover a $\sigma$-model for this configuration as $\omega$ has no poles at finite distance.
    
    \subsubsection{II: An Undefined Configuration} 

    \begin{table}[h!]
        \centering
        \def\arraystretch{1.5}
        \begin{tabular}{c|c|c}
             $\omega$ & \multicolumn{2}{|c}{P-defects} \\ 
            \hline
             \multirow{3}{*}{$\dfrac{dz}{z}$} & Type & Position in $\mathbb{C}P^1 $ \\ 
             \cline{2-3}
             & Chiral & $0$  \\
             \cline{2-3}
             & Chiral & $\infty$
        \end{tabular} 
        \caption{The defect configuration in example II.}
        \label{table:3}
    \end{table}

    The defect configuration for this example is given in Table \ref{table:3}. 
    Again, since $\omega$ has no zeros, we do not introduce any Z-defects, and the solution \eqref{Lax solution} for $\mathcal A$ simplifies to: 
    \begin{equation}
        \lax{A} = \lax{A}_{+}^{c}(x^{+},x^{-}) dx^{+} + \lax{A}_{-}^{c}(x^{+},x^{-}) dx^{-} \fp
    \end{equation}
    Using equation \eqref{bc equation}, the two boundary conditions, $g_{0} = g$ and $g_{\infty}=1$ we find the simultaneous equations: 
    \begin{equation}
        \mathcal A_{-}^c = g^{-1} \partial_{-} g \ca \quad \mathcal A_{-}^c = 0 \ca 
    \end{equation}
    which leaves $\mathcal A_{+}$ undetermined and $\mathcal A_{-}$ constrained to be a function of $x^{+}$ only. 
    Once substituted into the \eqref{unified sigma model action}, the latter condition produces a vanishing $\sigma$-model action. 
    This shows we must be careful to choose boundary conditions which completely fix the field configuration.
    
    \subsubsection{III: The Wess-Zumino-Witten Model} 

    \begin{table}[h!]
        \centering
        \def\arraystretch{1.5}
        \begin{tabular}{c|c|c}
             $\omega$ & \multicolumn{2}{|c}{P-defects} \\ 
            \hline
             \multirow{3}{*}{$k \dfrac{dz}{z}$} & Type & Position in $\mathbb{C}P^1 $ \\ 
             \cline{2-3}
             & Chiral & $0$  \\
             \cline{2-3}
             & Anti-chiral & $\infty$
        \end{tabular} 
        \caption{The defect configuration in example III.}
        \label{table:4}
    \end{table} 

    The defect configuration for this example is given in Table \ref{table:4}. 
    As in the previous two examples, $\omega$ has no zeros, and therefore there are no Z-defects. 
    Hence, the solution \eqref{Lax solution} for $\mathcal A$ simplifies to: 
    \begin{equation}
        \lax{A} = \lax{A}_{+}^{c}(x^{+},x^{-}) dx^{+} + \lax{A}_{-}^{c}(x^{+},x^{-}) dx^{-} \fp
    \end{equation}
    Substituting the P-defect boundary conditions in equation \eqref{bc equation}, we find the components of the Lax connection are: 
    \begin{equation}
        \mathcal A_{-}^{c} = g^{-1} \partial_{-} g \ca \quad \mathcal A_{+}^{c} = 0 \ca 
    \end{equation} 
    This is an example of a Lax connection without a spectral parameter. 
    Upon substituting this solution into \eqref{unified sigma model action}, one finds the Wess-Zumino-Witten model:
    \begin{equation}
		S_{\text{WZW}}(g) = \frac{k}{4 \pi} \int_{\Sigma_{0}} d^{2} x \tr{ g^{-1} \partial_{+} g}{ g^{-1} \partial_{-} g} - \frac{k}{24 \pi} \int_{M_0} \tr{\tilde{g}^{-1} d \tilde{g}}{[\tilde{g}^{-1} d \tilde{g},\tilde{g}^{-1} d \tilde{g}]} \fp \label{WZW mod 2}
	\end{equation} 
    Using the flatness of the Lax connection, we do indeed find the WZW model's equations of motion: 
    \begin{equation}
        \partial_{+} (g^{-1} \partial_- g ) = g^{-1} \partial_- ( \partial_+ g g^{-1} ) g = 0 \fp
    \end{equation}
    
	\subsubsection{IV: The Principal Chiral Model} 

    \label{PCM section}

    \begin{table}[h!]
        \centering
        \def\arraystretch{1.5}
        \begin{tabular}{c|c|c|c|c}
             $\omega$ & \multicolumn{2}{|c|}{P-defects} & \multicolumn{2}{|c}{Z-defects} \\ 
            \hline
             \multirow{3}{*}{$\dfrac{1}{2}\dfrac{z^2-\lambda^2}{(z-k)^2} dz$} & Type & Position in $\mathbb{C}P^1 $ & Type & Position in $\mathbb{C} P^1$ \\ 
             \cline{2-5}
             & Dirichlet & $k$ & $(1,0)$ & $\lambda$ \\
             \cline{2-5}
             & Dirichlet & $\infty$ & $(0,1)$ & $-\lambda$ 
        \end{tabular} 
        \caption{The defect configuration in example IV.}
        \label{table:5}
    \end{table} 

    In this final example, we use the defect configuration given in Table \ref{table:5}. Its primary purpose is to illustrate an example where $\omega$ has zeros. 
    Our starting point for this construction is to consider: 
    \begin{equation}
        \omega = \frac{1}{2}\frac{(z-\zeta_1)(z-\zeta_2)}{(z-k)^2} dz \ca 
    \end{equation}
    whose zeros and poles introduce four marked points on $\mathbb{C}P^1$: $\zeta_{1,2}$, $k$ and $\infty$. 
    In general, one can perform a modular transformation to fix three of these points. 
    In this case, we have chosen to keep the pole at $\infty$, and then send the two zeros to\footnote{We could have used this transformation to set $\lambda = 1$, but defer doing so as we wish to discuss the $\lambda \rightarrow 0$ limit shortly.} $\lambda$ and $-\lambda$, which produces the $\omega$ given in Table \ref{table:5}. 
    The overall factor of a half in $\omega$ is there as a matter of convention.
    
    At the zeros $\pm \lambda$ we introduce $(1,0)$ and $(0,1)$ Z-defects. 
    Thus, we find \eqref{Lax solution} becomes: 
    \begin{equation}
        \mathcal A = \left( \mathcal A^{c}_+ + \frac{\mathcal A_{+}^{(\lambda)}}{z-\lambda} \right) dx^+  +  \left( \mathcal A^{c}_- + \frac{\mathcal A_{-}^{(-\lambda)}}{z+\lambda} \right) dx^- \fp
    \end{equation} 
    Using \eqref{bc equation} and the two boundary conditions at $z=k$ and $\infty$, we find the simultaneous equations: 
    \begin{align}
        \mathcal A^{c}_+ + \frac{\mathcal A_{+}^{(\lambda)}}{k-\lambda} &= g^{-1} \partial_{+} g \ca & \mathcal A^{c}_- + \frac{\mathcal A_{-}^{(-\lambda)}}{k+\lambda} &= g^{-1} \partial_{-} g \ca \\
        \mathcal A_{+}^c &=0 & \mathcal A_{-}^c &= 0 \ca
    \end{align}
    where we've set $g_0 = g$, as before. 
    Together, these yield the Lax connection of the principal chiral model with a Wess-Zumino term: 
    \begin{equation}
        \mathcal A = \frac{(k-\lambda) g^{-1} \partial_{+} g}{z-\lambda} d x^+ +\frac{(k+\lambda)g^{-1} \partial_- g}{z+\lambda} dx^- \ca
    \end{equation}
	which we can verify by substituting the above into the unified $\sigma$-model action to find: 
    \begin{equation}
       S_{\text{PCMWZ}} (g) = \frac{\lambda}{4 \pi} \int_{\Sigma_0} d^2 x \tr{g^{-1} \partial_{+} g}{g^{-1} \partial_{-} g} - \frac{k}{24 \pi} \int_{M_0} \tr{\tilde{g}^{-1} d \tilde{g}}{[\tilde{g}^{-1} d \tilde{g},\tilde{g}^{-1} d \tilde{g}]} \fp
    \end{equation} 
    The flatness of $\mathcal A$ produces the equations of motion: 
    \begin{equation}
         \partial_+ J_- + \partial_- J_+ = 0 \ca \quad \partial_+ J_- - \partial_- J_+ + [J_{+} ,J_{-}] \ca
    \end{equation}
    where we've defined the currents $J_{\pm} = \lambda (\lambda \mp k) g^{-1} \partial_{\pm} g$. 
    
    As a final remark, we consider the following three limits of $\omega$'s parameters:
    \begin{itemize}
        \item[$(a)$] $k \rightarrow 0$, which produces the standard principal chiral model; 
        \item[$(b)$] $k \rightarrow \lambda$, in which the Dirichlet defect at $k$ and the $(0,1)$ defect at $\lambda$ collide and cancel, leaving behind a chiral defect and the Wess-Zumino-Witten action;
        \item[$(c)$] and finally, $\lambda \rightarrow 0$, which fuse the $(1,0)$ and $(0,1)$ Z-defects to produce an $(1,1)$ defect and the (topological) Wess-Zumino model. 
    \end{itemize}

\section{Doubled Four-Dimensional Chern-Simons}

	\label{doubled 4dcs section} 

    During the last few sections, we've developed the classical theory of 4d CS and shown how it's used to construct IMs. 
    Most pertinently, we found that 4d CS describes integrable $\sigma$-models because the field $A$ is gauge equivalent to a Lax connection $\mathcal A$. 
    It of course follows from the preceding analysis that a collection of 4d CS theories will describe a set of integrable models, each with its own associated Lax connection. 
    Our goal in the next couple of sections is to ask whether multiple 4d CS theories can be coupled together such that they still describe an IM. 
    We find that this is indeed possible, and that the resulting models have a gauge symmetry. 
    Unlike above, we restrict $\omega$ to have simple poles only and simplify our notation to $k_q = k_q^{(0)}$.
    This will guarantee that the action is regular and enable the use of the archipelago conditions when reducing the action. 

    We begin by defining the ``doubled 4d CS'' theory and describe its basic properties. 
    Next, we derive the equations of motion and introduce ``gauged'' P-defects with their associated boundary conditions. 
    Finally, we will prove that the theory is gauge invariant.

    \subsection{The Doubled Action} 

    \label{section: the doubled action}

    We will focus on the simplest version of this construction, which has two gauge fields $A$ and $B$. 
    The gauge groups of these fields are again assumed to be complex semi-simple and will be denoted by $G$ and $H \subseteq G$, respectively. 
    Their Lie algebras are $\mathfrak g$ and $\mathfrak h \subseteq \mathfrak g$, with the latter determined by an embedding into the former $\pi : \mathfrak h \hookrightarrow \mathfrak{g}$. 
    We restrict ourselves such that the subgroup $H$ forms a left coset $G / H$ that is a reductive homogeneous space. 
    This means we can decompose $\mathfrak g$ into $\mathfrak{g} =\mathfrak{f} \oplus \mathfrak{h}$ such that the subspaces $\mathfrak h$ and $\mathfrak f$ satisfy:
	\begin{equation}
		[\mathfrak{h},\mathfrak{h}] \subset \mathfrak{h} \ca \quad [\mathfrak{h},\mathfrak{f}] \subset \mathfrak{f} \ca \quad \tr{\mathfrak{h}}{ \mathfrak{f}}_{\mathfrak{g}} = 0 \fp \label{red. hom. def}
	\end{equation}
	Since $\mathfrak{f}$ is orthogonal to $\mathfrak{h}$ by the final expression in the above, we call it the orthogonal complement. 
    Finally, to simplify several expressions that appear in the following, we introduce the notation $X^{\mathfrak{h}}$ which indicates the projection of  $X \in \mathfrak g$ to the embedded copy of $\mathfrak h$. 
    More generally, if an expression is projected onto $\mathfrak{h}$ then we place $\proh$ to the right.
    Likewise, a superscript $\mathfrak{f}$ and $\prof$ denote the projection onto $\mathfrak{f}$.
    
    To construct the 4d CS action associated to each of these fields, we defined a bilinear form $\tr{\cdot}{\cdot}_{\mathfrak{g}}$ on $\mathfrak{g}$ as before, and note that the embedding $\pi$ induces $\tr{\cdot}{\cdot}_{\mathfrak{h}}$ on $\mathfrak{h}$ via:
    \begin{equation}
		\iota \tr{a}{b}_{\mathfrak{h}} = \tr{\pi(a)}{\pi(b)}_{\mathfrak{g}} \ca \label{relation btwn traces}
	\end{equation}
	where $\iota$ is called the index of embedding \cite{Fuchs:1997jv}. 
    We will again assume that $\mathfrak{g}$ is in the adjoint representation $R_{\text{ad}}$, which in turn induces a representation $R_{ad} \circ \pi$ of $\mathfrak{h}$. 
    Using these forms and the two fields $A \in \mathfrak{g}$ and $B \in \mathfrak{h}$, we define the \textit{doubled 4d CS} theory to be the difference of two 4d CS actions, one for each field, plus a new boundary term which couples $A$ and $B$ together:
	\begin{align}
		S_{\text{Dbld}}(A,B) &= S_{\text{4dCS}}(A) - S_{\text{4dCS}}(B) + S_{\text{bdry}}(A,B) \nonumber \\
		&= \frac{i}{8 \pi^2} \int_{M} \omega \wedge \tr{A}{ dA + \tfrac{1}{3} [A, A]}_{\mathfrak{g}} - \frac{i \iota}{8 \pi^2} \int_{M} \omega \wedge \tr{B}{ d B + \tfrac{1}{3} [B, B]}_{\mathfrak{h}} \nonumber \\
        & \ \ \ \ \ \ \ \ \ \ \ \ \ \ \ \ \ \ \ \ \ \ \ \ \ \ \ - \frac{i \iota}{8 \pi^2} \int_{M} \omega \wedge \cdel \tr{A^{\mathfrak{h}}}{ B}_{\mathfrak{h}} \ca \label{doubled action}
	\end{align} 
    where again $M = \Sigma \times \mathbb{C} P^1$. 
	We will often refer to this action as the doubled action/theory for short. 
    The final term in the above action collapses to a boundary term via equation \eqref{integral over poles} and thus only modifies the defect equations of motion. 

    Later in this section, we will prove that the doubled action is gauge invariant using the relation \eqref{relation btwn traces} and the requirement that $G/H$ be a reductive homogeneous space. 
    For the moment, we will assume that this is indeed the case and impose \eqref{relation btwn traces} to simplify our expressions and notation. 
    Upon doing this, we treat $B$ as a gauge field valued in $\mathfrak{g}$, whose components in $\mathfrak{f}$ vanish and drop the projection of $A$ in the boundary term since $\tr{A^{\mathfrak{f}}}{B}$ vanishes by \eqref{red. hom. def}. 
    When we come to discussing gauge invariance, we will reintroduce the two bilinear forms and show that the two aforementioned requirements are necessary. 

    Just as the standard action had an unusual gauge symmetry, so too does the doubled one given by: 
    \begin{equation}
		A_{z} \rightarrow A_{z} + \chi_{z} \ca \qquad B_{z} \rightarrow B_{z} + \xi_{z} \ca \label{2.5}
	\end{equation}
	where $\chi_{z}$ and $\xi_{z}$ are arbitrary functions valued in $\mathfrak{g}$ and $\mathfrak{h}$, respectively. 
    Again, all field configurations of $A_z$ and $B_z$ are gauge equivalent, allowing us to fix the gauges $A_{\bar z} = B_{\bar z} = 0$, simplifying $A$ and $B$ to:
	\begin{equation}
		A = A_{+} dx^{+} + A_{-} dx^{-} + A_{\bar{z}} d \bar{z} \ca \qquad B= B_{+} dx^{+} + B_{-} dx^{-} + B_{\bar{z}} d \bar{z} \fp
	\end{equation}
    The action is also semi-topological. 

    \subsection{The Equations of Motion} 

    In this section, we derive the doubled theory's bulk and defect equations of motion, solve the former in the Lax gauge, and finally, define the gauged P-defects. 

    As usual, the equations of motion are found from the action by performing its variation. 
    In this case, we find:
	\begin{align}
		\delta S_{\text{Dbld}}(A,B) = \frac{i}{4 \pi^2} \int_{M} \omega \wedge & \tr{F(A)}{ \delta A } - \frac{i}{4 \pi^2} \int_{M} \omega \wedge \tr{ F(B)}{\delta B} \label{doubled action var.} \\ &- \frac{i}{8 \pi^2} \int_{M} \omega \wedge \cdel \tr{A-B}{\delta A + \delta B} \;. \nonumber
	\end{align} 
    As was done earlier, we require that each of these terms vanish separately. 
    This implies that the \textit{doubled bulk equations of motion} are:
	\begin{equation}
		\omega \wedge F(A) = 0 \ca \qquad \omega \wedge F(B) = 0 \ca \label{bulk deom}
	\end{equation}
    while the second term produces:
	\begin{equation}
		\frac{i}{8 \pi^2} \int_{M} \omega \wedge \cdel \tr{A-B}{\delta A + \delta B} = 0 \ca \label{doubled beom}
	\end{equation} 
    which after using \eqref{integral over poles} collapses to: 
    \begin{equation}
        \frac{1}{4 \pi} \sum_{q \in P} k_q \int_{\Sigma_q} \tr{A_\Sigma - B_\Sigma}{\delta A_\Sigma + \delta B_\Sigma} = 0 \fp \label{eq: doubled beom}
    \end{equation}

    \subsubsection{More Lax Connections} 

    \label{section: more lax connections}

    Let's consider solutions to the bulk equations of motion in the Lax gauge defined by $A_{\bar z} = B_{\bar z} = 0$, $A_{\Sigma} = \mathcal A$ and $B_{\Sigma} = \mathcal B$. 
    As in section \ref{section:Lax connection}, when this gauge choice is, made the bulk equations of motion reduce to: 
    \begin{align}
        \omega \wedge \cdel \mathcal A &= 0 \ca & d_{\Sigma} \mathcal A_{\Sigma} + \frac{1}{2} [\mathcal A_\Sigma , \mathcal A_\Sigma] &= 0 \ca \\ 
        \omega \wedge \cdel \mathcal B &= 0 \ca & d_{\Sigma} \mathcal B_{\Sigma} + \frac{1}{2} [\mathcal B_\Sigma , \mathcal B_\Sigma] &= 0 \ca
    \end{align} 
    and we call $\mathcal A$ and $\mathcal B$ Lax connections because they satisfy the three required properties given in section \ref{section:Lax connection}. 
    Their associated monodromy matrices are: 
    \begin{equation}
        U(z) = \text{P} \exp \int_{\gamma} \mathcal A \ca \qquad V(z) = \text{P} \exp \int_\gamma \mathcal B \ca
    \end{equation}
    where $\gamma$ is curve spanning a timeslice of $\Sigma$. 

    The Z-defects of the doubled theory are again defined using the regularity conditions given in equation \eqref{eq: Z-defect definition}, but with $B$ substituted for $A$ if we want to introduce a singularity in $B$. 
    We will use a subscript $A$ or $B$ to indicate which field the defect applies to, \textit{e.g.} $(1,0)_A$ and $(0,1)_B$-defects introduce poles in $A_{+}$ and $B_{-}$, respectively. 

    The solution to $\omega \wedge \cdel \mathcal A = 0$ in the presence of a set of Z-defects is again given by equation \eqref{Lax solution}. 
    We also solve $\omega \wedge \cdel \mathcal B = 0$ in exactly the same way, although the notation we use is slightly different: 
    \begin{equation}
        \mathcal B_+ = \mathcal B^c_{+} + \sum_{\zeta \in Z} \sum_{l=0}^{m_{\zeta}^+-1} \frac{\Psi_\zeta^{(l)}}{(z-\zeta)^{l+1}} \ca \quad \mathcal B_- = \mathcal B^c_{-} + \sum_{\zeta \in Z} \sum_{l=0}^{m_{\zeta}^- -1} \frac{\Phi_\zeta^{(l)}}{(z-\zeta)^{l+1}}  \ca \label{solution for Lax B}
    \end{equation} 
    in which we again assume $m_\zeta^+ + m_\zeta^- \leq m_\zeta$. 
    Our motivation for making this change is that the coefficients $\Psi_\zeta^{(l)}$ and $\Phi_{\zeta}^{(l)}$ are scalar fields within the $\sigma$-models considered later.

	\subsubsection{Boundary Conditions and Gauged P-Defects}
	
	\label{section:boundary conditions} 

    Let's turn now to equation \eqref{eq: doubled beom} and a discussion of its solutions. 
    We will again restrict ourselves to the case where each of the integrands in the sum over the poles vanishes independently and thus find the \textit{doubled defects equations of motion:} 
    \begin{equation}
        \epsilon^{ij} \tr{A_i \pp- B_i \pp}{\delta A_j \pp + \delta B_j \pp} = 0 \ca \label{eq: doubled defect eom}
    \end{equation} 
    where $\epsilon^{+-} = 1$. 
    As we shall see momentarily, the boundary conditions that solve these equations relate $A$ and $B$ to each other in a way that preserves an $H \subseteq G$ gauge symmetry on the associated defect. 
    We therefore call these ``gauged'' P-defects. 

    We can further simplify \eqref{eq: doubled defect eom} by using the decomposition $\mathfrak g = \mathfrak h \oplus \mathfrak f$ and the fact that $\mathfrak h$ and $\mathfrak f$ are orthogonal to each other. 
    Upon doing this, we find two separate equations: 
    \begin{align}
        \epsilon^{ij} \tr{A_{j}^{\mathfrak{f}} \pp}{\delta A_{k}^{\mathfrak{f}} \pp }  &= 0 \ca \label{2.12} \\ 
        \epsilon^{ij} \tr{A_{j}^{\mathfrak{h}} \pp - B_{j} \pp}{\delta A_{k}^{\mathfrak{h}}\pp + \delta B_{k}\pp} &= 0 \label{2.13} \fp
	\end{align} 
    When solving these equations, we restrict ourselves to considering gauged analogues of the chiral and anti-chiral conditions discussed earlier. 
    These solutions can be found in the second and third columns of Table \ref{table:6}, while the fourth column contains the constraint imposed upon the generator $(u,v) \in G \times H$ of the gauge transformation: 
    \begin{equation}
		A \rightarrow {}^{u} A := u(d+A)u^{-1} \ca \qquad B \rightarrow {}^{v} B := v(d+B)v^{-1} \ca \label{2.14}
	\end{equation} 
    by the requirement that the boundary condition be preserved. 
    The group $K$ in this column is given by the intersection:
    \begin{equation}
        K = C(\mathfrak{h}) \cap N(\mathfrak{f}) \ca
    \end{equation} 
    where $C(\mathfrak{h})$ and $N(\mathfrak{f})$ are, respectively, the centraliser of $\mathfrak{h}$ and normaliser of $\mathfrak f$ within $G$. 
    These constraints will be essential when proving that the doubled action is gauge invariant, and will be derived in the remainder of this section. 
    Note that we assume a boundary condition $A_{i}^\mathfrak{f} \pp = 0$ implies $\delta A_{i}^\mathfrak f \pp = 0$ to achieve the equality in equation \eqref{2.12}. 

    \begin{table}[h!]
        \def\arraystretch{1.5}
        \centering
        \begin{tabular}{c|c|c|c}
            \multirow{2}{*}{Defect Type} & \multirow{2}{30mm}{Genuine Boundary \\ \ \ \ \ \  Conditions} & \multirow{2}{*}{Gauging Constraint} & \multirow{2}{30mm}{\ Constraint on \\ $y_q = u_q^{-1} v_q \in K$} \\ 
            & & & \\
            \hline 
            Gauged Chiral & $A_{-} \pp = B_{-} \pp $ & $A_+^\mathfrak h \pp = B_+ \pp$ & $\partial_- y_q = y^{-1}_q \partial_{+} y_q |^{\mathfrak h} = 0$ \\
            \hline
            Gauged Anti-Chiral & $A_{+} \pp = B_{+} \pp $ & $A_-^\mathfrak h \pp = B_- \pp$ & $\partial_+ y_q = y^{-1}_q \partial_{-} y_q |^{\mathfrak h} = 0$
        \end{tabular}
        \caption{The gauged chiral and anti-chiral boundary conditions and their associated constraints on gauge transformations.}
        \label{table:6}
    \end{table}

    The reader is likely to have noticed that we've split the boundary conditions into two subclasses. 
    Our reason for doing so arises from the role the two classes play when constructing $\sigma$-models from the doubled theory. 
    The first class, called ``genuine boundary conditions'', are imposed when solving the bulk equations of motion $\omega F_{\bar z -} (A) = \omega F_{\bar z -} (B) = 0$, and allow us to specify $\mathcal A$ in terms of $\mathcal B$ and a series of group elements $g_q$ associated to the defects. 
    The second class, which we call ``gauging constraints'', are not imposed during this step and instead are equations of motion within the resulting $\sigma$-model. 
    We have chosen this name because they constrain a Kac-Moody current, and are a result of the gauge symmetry. 


    Let's derive the constraints on gauge transformations given in Table \ref{table:6}. 
    We start with the gauge chiral conditions: 
    \begin{gather}
		A_{-}^{\mathfrak{f}} \pp = 0 \ca \qquad A_{\pm}^\mathfrak{h} \pp - B_{\pm} \pp = 0 \fp 
    \end{gather} 
    where the requirement that they are preserved by gauge transformations produces the following constraints on $u$ and $v$: 
    \begin{align}
		({}^{u} A_{-})^{\mathfrak{f}} \pp = (u_q B_{-}\pp u_q^{-1} + u_q \partial_{-}u_q^{-1})\prof = 0 \ca \label{2.19} \\
        ({}^{u} A_{\pm})^{\mathfrak{h}} \pp - {}^{v} B_{\pm}\pp =  (u_q A_{\pm} \pp u^{-1}_q + u_q \partial_{\pm} u^{-1}_q)\proh - (v_qB_{\pm}\pp v^{-1}_q + v_q\partial_{\pm} v_q^{-1}) = 0 \fp \label{eq: gcbc 2nd constraint}
	\end{align} 
    Using the change of variables $u=v y^{-1}$, we can recast these equations as constraints on $y = u^{-1} v \in G$. 
    
    The substitution of $u = v y^{-1}$ into equation \eqref{2.19} leads to: 
    \begin{equation}
        (v_q (y^{-1}_q B_{-} \pp y_q + y^{-1}_q \partial_{-} y_q) v^{-1}_q + v_q \partial_{-} v^{-1}_q)\prof = 0 \fp
    \end{equation} 
    If we decompose $y^{-1} B_{-} y + y^{-1} \partial_{-} y$ into its $\mathfrak{f}$ and $\mathfrak{h}$ parts, use $v \mathfrak{f} v^{-1} = \mathfrak{f}$ and $v \mathfrak{h} v^{-1} = \mathfrak{h}$ (as $v \in H$ and $[\mathfrak{h},\mathfrak{f}] \subset \mathfrak{f}$), and project into $\mathfrak{f}$ 
    this becomes: 
    \begin{gather}
		(y^{-1}_q B_{-} \pp y_q + y^{-1}_q \partial_{-} y_q) \prof = 0 \fp \label{2.17} 
    \end{gather}
    Equation \eqref{eq: gcbc 2nd constraint} also simplifies under the subsitution of $u=vy^{-1}$ into: 
    \begin{equation}
		v_q(y_q^{-1}A_{\pm}\pp y_q+y_q^{-1}\partial_{\pm}y_q-B_{\pm}\pp)v_q^{-1} \proh = 0 \label{2.21} \ca
	\end{equation}
    where again use $v \mathfrak{f} v^{-1} = \mathfrak{f}$ and $v \mathfrak{h} v^{-1} = \mathfrak{h}$. Hence, we find \eqref{eq: gcbc 2nd constraint} becomes: 
    \begin{equation}
		(y_q^{-1} A_{\pm} \pp y_q + y^{-1}_q \partial_{\pm} y_q) \proh = B_{\pm} \pp \fp \label{eq: 4.24}
	\end{equation} 

    When discussing the gauge invariance of an action, one considers arbitrary field configurations satisfying the boundary conditions, not just those which are on-shell. 
    For this reason, equations \eqref{2.17} and \eqref{eq: 4.24} must hold for every $A$ and $B$. 
    Therefore, \eqref{2.17} implies the conditions $y \pp \in C(\mathfrak{h})$, and $\partial_{-} y \pp = 0$.
    Using these conditions, equation \eqref{eq: 4.24} reduces to $(y^{-1} A_{+}|^\mathfrak{f} y + y^{-1} \partial_{+} y)\proh_{\textbf{\textit{q}}} = 0$ which is satisfied if $y^{-1} \partial_{+} y\proh_{\textbf{\textit{q}}} = 0$ and $y \pp \in N(\mathfrak{f})$.

    Following a similar argument for the gauged anti-chiral boundary conditions: 
    \begin{gather}
		A_{+}^{\mathfrak{f}} \pp = 0 \ca \qquad A_{\pm}^{\mathfrak{h}} \pp - B_{\pm} \pp = 0 \ca 
    \end{gather} 
    we find the constraints: 
    \begin{gather}
		(y_q B_{+}\pp y_q^{-1} + y_q\partial_{+}y^{-1}_q)\prof = 0 \ca \qquad (y_q^{-1}A_{\pm}\pp y_q+y_q^{-1}\partial_{\pm}y_q)\proh= B_{\pm} \pp \fp 
	\end{gather}
    These are solved in the same way as equations \eqref{2.17} and \eqref{eq: 4.24} and produce the constraints given in Table \ref{table:6}. 
	
	\subsection{Gauge Invariance}
	
	In this section, we prove that the doubled action \eqref{doubled action} is gauge invariant for field configurations that satisfy the gauged chiral/anti-chiral conditions introduced in previously. 
    As was mentioned in section \ref{section: the doubled action}, this will require that the coset $G/H$ be a reductive homogeneous space, and that the bilinear forms of $\mathfrak g$ and $\mathfrak h$, $\tr{\cdot}{\cdot}_{\mathfrak{g}}$ and $\tr{\cdot}{\cdot}_{\mathfrak{h}}$, satisfy the relation \eqref{relation btwn traces}. 
    To ensure that the second of these requirements is clear, we work with the initial expression for the action given in equation \eqref{doubled action}.

	Under the gauge transformations $A \rightarrow {}^u A$ and $B \rightarrow {}^v B$ the doubled action transforms as:
	\begin{equation}
		S_{\text{Dbld}} (A,B) \rightarrow S_{\text{Dbld}} ({}^{u} A,{}^{v} B) = S_{\text{4dCS}}({}^{u} A) - S_{\text{4dCS}}({}^{v} B) + S_{\text{bdry}}({}^{u} A,{}^{v} B) \ca \label{doubled action gauge trans.}
	\end{equation}
	where $S_{\text{4dCS}} ({}^u A)$ is given in equation \eqref{cwy gauge transformed}, $S_{\text{4dCS}} ({}^v B)$ is of the same form, but with $B$ and $v$ substituted for $A$ and $u$, and:
    \begin{equation}
		S_{\text{bdry}}({}^{u} A,{}^{v} B) = - \frac{i}{8 \pi^2} \int_{M} \omega \wedge \cdel \tr{uAu^{-1} -duu^{-1} }{v B v^{-1} -dv v^{-1}}_{\mathfrak{h}} \fp \label{gauge boundary term}
	\end{equation}
	In these expressions, we have sent a total derivative in $d_{\Sigma}$ to zero by requiring $A$, $B$, $u$ and $v$ die off sufficiently fast enough at infinity. 

    To show the action is gauge invariant, we substitute in $u = v y^{-1}$, use equation \eqref{relation btwn traces}, the Polyakov-Wiegmann identity \cite{Polyakov:1984et}:
	\begin{align}
		\tr{u^{-1}du}{[u^{-1}du,u^{-1}du]} -& \tr{v^{-1}dv}{[v^{-1}dv,v^{-1}dv]} \nonumber \\
        &= \tr{ydy^{-1}}{[ydy^{-1},ydy^{-1}]} + 6 d \tr{vdv^{-1}}{du u^{-1}} \fp \label{General polyakov-wiegmann}
	\end{align}
    After doing this, the two Wess-Zumino terms in \eqref{doubled action gauge trans.} can be written as: 
    \begin{equation}
        S_{\text{4dWZ}}(u)-S_{\text{4dWZ}}(v) = S_{\text{4dWZ}}(y^{-1}) + \frac{i}{8 \pi^2} \int_M \omega \wedge \tr{v d v^{-1}}{d u u^{-1}} \fp 
    \end{equation}
    and after cancelling several terms, we find:
	\begin{align}
		S_{\text{Dbld}} ({}^{u} A,{}^{v} B) = S_{\text{4dCS}}(A) &- S_{\text{4dCS}} (B) + S_{\text{4dWZ}} (y^{-1}) \nn \\
		& + \frac{1}{2 \pi \hbar} \int_{M} \omega \wedge \cdel \left( \tr{ydy^{-1}}{A} - \tr{y^{-1}Ay + y^{-1} dy}{B} \right) \fp \label{2.40}
	\end{align} 
    
	In the previous section, we used the conditions required of $\mathfrak h$ and $\mathfrak f$ such that $G/H$ is reductive homogeneous to derive a series of constraints upon $y$. 
    The most pertinent of these is the requirement that $y \in K$, which once imposed upon the above equation simplifies the final term to $S_{\text{Bdry}}(A,B)$ and leads us to: 
    \begin{equation}
        S_{\text{Dbld}} ({}^{u} A,{}^{v} B) = S_{\text{Dbld}}(A,B) + \frac{i}{8 \pi^2} \int_{M} \omega \wedge \cdel \tr{ydy^{-1}}{A} + S_{\text{4dWZ}} (y^{-1}) \fp
    \end{equation}
    From here, the proof of gauge invariance proceeds in exactly the same way as in section \ref{section:cwy-gauge-invariance}. 
    That is, using \eqref{integral over poles}, we can collapse the second term on the right-hand side down to a sum over poles and then show the result vanishes after using: 
    \begin{equation}
        \partial_- y_q =0 \ca \qquad \partial_+ y_q = 0 \ca
    \end{equation} 
    for gauged chiral and anti-chiral conditions, respectively. 
    For the Wess-Zumino term, we again suppose $\Sigma = \mathbb{R}^2$, which implies the Wess-Zumino three-form is closed and exact. 
    This means we can collapse the Wess-Zumino term down to a sum over poles by introducing group coordinates on $K$ and using equation \eqref{integral over poles}. 
    The result of this calculation will vanish once the above conditions are imposed. 
	
\section{The Unified Gauged Sigma Model}
	
	\label{section:UGSM} 

    In this section, we reduce the doubled action down to a ``unified gauged $\sigma$-model'' located on the gauged P-defects inserted at $\omega$ poles. 
    We will do this by essentially following the DLMV argument that was reviewed in section \ref{section: Integrable Sigma Models}. 
    Our starting point is a simultaneous change of variables and partial gauge fixing in which we go from $(A,B)$ to the group elements $(\hat \sigma,\hat \rho)$ and Lax connections $(\mathcal A, \mathcal B)$. 
    As before, this is done via the expressions: 
    \begin{align}
        \widebar{A} = \hat{\sigma} \cdel \hat \sigma^{-1} \ca \quad A_{\Sigma} = \hat \sigma d_{\Sigma} \hat \sigma^{-1} + \hat \sigma \mathcal A \hat \sigma^{-1} \\ 
        \widebar{B} = \hat{\rho} \cdel \hat \rho^{-1} \ca \quad B_{\Sigma} = \hat \rho d_{\Sigma} \hat \rho^{-1} + \hat \rho \mathcal B \hat \rho^{-1} \fp
    \end{align}
    The second step in this construction is to define a set of archipelago conditions for a pair of group elements $(\tilde g, \tilde h)$, and show that we can always access the corresponding archipelago gauge by performing an appropriate gauge transformation. 
    Finally, we set $(\hat \sigma, \hat \rho) = (\tilde g , \tilde h)$ in the above equations, substitute the result into the doubled action and use the archipelago conditions to find the unified gauge $\sigma$-model. 

    \subsection{$\Sigma$-Model Fields and The Archipelago Conditions revisited}

    \label{gfge} 

    As in section \ref{vicedo's construction}, given the fields $A_{\bar z}$ and $B_{\bar z}$ we can construct a pair of group elements $(\hat \sigma, \hat \rho) : \Sigma \times \mathbb{C} P^1 \rightarrow G \times H$ defined by:  
    \begin{equation}
		A_{\bar z} = \hat{\sigma} \cdel \hat{\sigma}^{-1} \ca \qquad B_{\bar z} = \hat{\rho} \cdel \hat{\rho}^{-1} \fp \label{4.1}
	\end{equation}
    These expressions are invariant under a right-redundancy transformation $\hat \sigma \rightarrow \hat \sigma k_g$ and $\hat \rho \rightarrow \hat \rho k_g$, where $k_{g,h}$ are functions of $x^\pm$ only, and thus define two equivalence classes, $\{ \hat \sigma \}$ and $\{ \hat \rho \}$. 
    We fix the right-redundancy by selecting from these classes the canonical pair of elements $(\hat{g},\hat{h})$, which have the defining property $\hat g (\infty) = \hat h (\infty) = 1$. 
    These can always be constructed by using equation \eqref{gpi def}. 

    To perform the next few steps in our analysis, we wish to work in the archipelago gauge of the doubled theory, which is defined by: 
    \begin{itemize}
        \item[(a)] $\hat g$ satisfies the archipelago conditions,
        \item[(b)] $\hat h$ is the identity.
    \end{itemize} 
    Doing so will enable us to reduce the doubled action down to a two-dimensional one. 
    
    To show that the archipelago gauge is accessible, we construct an admissible gauge transformation that puts any field configuration into this gauge. 
    If $A$ and $B$ define group elements $\hat g$ and $\hat h$, then this is equivalent to finding gauge transformations\footnote{Not to be confused with the monodromy matrices discussed earlier.} $U\in G$ and $V\in H$ consistent with the defect boundary conditions such that $\tilde g = U \hat g U_{\infty}^{-1}$ satisfies (a) and $\tilde h = V \hat{h} V_{\infty}^{-1}$ satisfies (b).
    We shall do this in two steps\footnote{Note that $\hat g^h \equiv\hat h^{-1}\hat g$ and does not denote conjugation by $h$, as is often meant by this notation}:
    \begin{equation}
        (\hat g,\hat h) \xrightarrow{(\hat h^{-1},\hat h^{-1}) } (\hat h^{-1} \hat g ,1) \equiv(\hat g^h,1) \xrightarrow{(\tilde{u},1)} (\tilde{u} \hat{g}^{h},1) \equiv(\tilde g,1) \ca 
    \end{equation}
    where we have used $\hat h_{\infty} = \hat u_\infty =1 $.

    In the first gauge transformation, we use the fact that gauge transformations of $B$ are unrestricted by the gauged chiral/anti-chiral boundary conditions we have defined above, and so simply choose $v = \hat{h}^{-1}$.
    This takes $B$ everywhere in $\remsph$ to:
	\begin{equation}
		B_{\bar{z}} = 0 \ca \qquad B_{\Sigma} = \lax{B} \fp \label{4.17}
	\end{equation}
    For this gauge transformation to be consistent, it must satisfy the conditions given in Table \ref{table:6}.
    These conditions are all satisfied if we fix $u=\hat{h}^{-1}$ since this implies that $y = u^{-1} v = 1$ everywhere on $\Sigma \times \remsph$. 

    To ensure the second step is clear, let's first discuss it for the gauge chiral conditions, and then move on to the gauge anti-chiral ones. 
    The required gauge transformation is $(u,v) = (\tilde{u},1)$ where $\tilde{u} = \tilde{g} (\hat{g}^h)^{-1}$ and it must satisfy: 
    \begin{equation}
        y \pp \in K \ca \quad \partial_{-} y \pp = 0 \ca \quad y^{-1} \partial_{+} y \proh_{\textbf{\textit{q}}} = 0 \fp
    \end{equation}
    This is certainly the case because archipelago condition $iii)$ implies that $y = \tilde{u}^{-1} = 1$ at a simple pole $q$; thus the above conditions are satisfied.
    The same argument works for anti-chiral boundary conditions if one swaps $+$ and $-$ in the equation above.

    As a result of the above, we are free to set: 
	\begin{equation}
		A = \tilde{g} d \tilde{g}^{-1} + \tilde{g} \lax{A} \tilde{g}^{-1} \ca \qquad B = \lax{B} \fp \label{4.31}
	\end{equation}

    Before moving to the next section, we wish to make one final observation concerning the symmetry transformations of $\tilde{g}$ and the Lax connection $\mathcal A$ and $\mathcal B$. 
    In section \ref{section: Symmetries of the Lax connection}, we argued that $A$ must be ignorant of our fixing of the right-redundancy as all elements of the class $\{ \hat \sigma \}$ produce the same field configurations for $A$. 
    The same argument is also true for $B$. 
    This implies that $\mathcal B$ is invariant under right-redundancy transformations because $\mathcal B = B$. 
    
    In the same section, we also argued that the residual gauge transformations of the field $A$ must preserve our fixing of the right-redundancy, and from this derived a transformation law for the $\tilde g$ and $\mathcal A$. 
    We can repeat this argument for the triple $(\tilde g, \mathcal A, \mathcal B)$ under the action of $A \rightarrow {}^u A$ and $B \rightarrow {}^v B$ to find: 
    \begin{equation}
        \tilde g \rightarrow {}^u \tilde g =u \tilde g u_{\infty}^{-1} \ca \quad \mathcal A \rightarrow {}^u \mathcal A  = u_{\infty} d_{\Sigma} u_{\infty}^{-1} + u_{\infty} \mathcal A u_{\infty}^{-1} \ca \quad \mathcal B \rightarrow {}^v \mathcal B = v d_{\Sigma} v^{-1} + v \mathcal B v^{-1} \ca
    \end{equation} 
    where we assume $u$ and $v$ preserve the archipelago conditions. 
    This requirement implies $v$ is independent of $z$ and $\bar z$. 
    If we compare the last of these three expressions with equation \eqref{solution for Lax B}, we can see that the coefficients $\mathcal B^c$, $\Psi_\zeta^{(l)}$ and $\Phi_{\zeta}^{(l)}$ transform as a gauge field and two adjoint scalars: 
    \begin{equation}
        \mathcal B^c \rightarrow v d_{\Sigma} v^{-1} + v \mathcal B^c v^{-1} \ca \qquad \Psi_\zeta^{(l)} \rightarrow v \Psi_\zeta^{(l)} v^{-1} \ca \qquad \Phi_\zeta^{(l)} \rightarrow v \Phi_\zeta^{(l)} v^{-1} \fp
    \end{equation}
    This offers us a higher-dimensional explanation for the origins of the gauge symmetry within $\sigma$-models considered below. 

 	\subsection{The Unified Gauged Sigma Model Action}

    \label{section: USGM} 

    In this subsection, we use the archipelago conditions to localise the doubled action \eqref{doubled action} down to the two-dimensional P-defects that sit at $\omega$'s poles. 
    We will find the unified $\sigma$-model action. 

    We start by assuming the gauge fields are in the archipelago gauge given by: 
    \begin{equation}
		A = \tilde{g} d \tilde{g}^{-1} + \tilde{g} \lax{A} \tilde{g}^{-1} \ca \qquad B = \lax{B} \fp
	\end{equation} 
    and substitute this expression into the doubled action. 
    These are again of the same form as a gauge transformation and thus the result of the substitution can be found from equation \eqref{2.40} with $y = \tilde{g}$. 
    We find:
	\begin{gather}
		S_{\text{Dbld}}(A,B) = \frac{i}{8 \pi^2} \int_{M} \omega \wedge \cdel \left( \tr{\tilde{g}^{-1} \dsig \tilde{g}}{\lax{A}} - \tr{- \dsig\tilde{g} \tilde{g}^{-1} + \tilde{g} \lax{A} \tilde{g}^{-1}}{\lax{B}} \right) + S_{\text{4dWZ}} (\tilde g) \label{7.33}
        \\+\frac{i}{8\pi^2} \int_{M} \omega \wedge \tr{\lax{A}}{\cdel \lax{A}} - \frac{i}{8 \pi^2} \int_{M} \omega \wedge \tr{\lax{B}}{\cdel \lax{B}}\nn \ca 
	\end{gather} 
    In the second step, we impose the equations of motion $\omega \wedge \cdel \mathcal A = \omega \wedge \cdel \mathcal B = 0$, which causes the final two terms to vanish, see section \ref{section: USM} for the explanation for why. 
    Finally, we use the archipelago conditions to integrate out any dependence on the angular coordinates of the disc $U_q$. 
    When doing this, the first and third terms produce the standard unified $\sigma$-model action, so we need only worry about the second term. 
    After using equation \eqref{integral over poles} to collapse the expression down to a sum over the poles of $\omega$ this becomes:
    \begin{align}
        - \frac{i}{8 \pi^2} \int_M \omega \wedge \cdel \tr{- \dsig\tilde{g} \tilde{g}^{-1} + \tilde{g} \lax{A} \tilde{g}^{-1}}{\lax{B}} &= - \frac{1}{4 \pi} \sum_{q \in P} \int  \text{Res}_{q}\left(\omega \wedge \tr{-\dsig g_{q} g_{q}^{-1} + g_{q} \lax{A} g_{q}^{-1}}{\lax{B}}\right)
    \end{align}
    Thus, the doubled action reduces to \textit{unified gauged $\sigma$-model}:
	\begin{align}
		S_{\text{UG$\Sigma$M}}(\tilde g ,\mathcal A, \mathcal B) := S_{\text{Dbld}}(A,B) = S_{\text{U$\Sigma$M}}(g,\lax{A}) - \frac{1}{4 \pi} \sum_{q \in P} \int_{\sigpi} \text{Res}_{q}\left(\omega \wedge \tr{-\dsig g_{q} g_{q}^{-1} + g_{q} \lax{A} g_{q}^{-1}}{\lax{B}}\right) \label{UGSM} \ca
	\end{align} 
    whose equations of motion are the flatness conditions: 
    \begin{equation}
        d_\Sigma \mathcal A + \frac{1}{2} [\mathcal A, \mathcal A] = 0 \ca \qquad d_\Sigma \mathcal B + \frac{1}{2} [\mathcal B, \mathcal B] \ca
    \end{equation} 
    and the gauge constraints of Table \ref{table:6}: 
    \begin{equation}
        (g \partial_i g^{-1} + g \mathcal A_i \pp g^{-1} )\proh = \mathcal B_i \pp \ca \label{eq: gauging constraints in terms of Laxes}
    \end{equation} 
    where $i$ is $+$ (resp. $-$) if the condition is gauged chiral (resp. anti-chiral). 

    \subsubsection{Example V: the Gauged Wess-Zumino-Witten model} 

    In this section, we construct the Lax connection and action of the gauged WZW model by taking $\omega = k \dfrac{dz}{z}$ with a gauge chiral and gauged anti-chiral P-defect.
    To do this, we use the genuine boundary conditions given in Table \ref{table:6} to construct a set of simultaneous equations, each of the form: 
    \begin{equation}
        \lax{A}_{i} \pp = g_{q}^{-1} \mathcal B_i \pp g_{q} + g_{q}^{-1} \partial_i g_{q} \ca \label{Lax from bc A} 
    \end{equation}
    where $i$ is $-$ (resp. $+$) if the condition is gauged chiral (resp. anti-chiral), and solve the resulting linear algebra problem to eliminate the coefficients of $\mathcal A$ in \eqref{Lax solution}. 
    This example is illustrative of a more general construction in which all boundary conditions of $A$ are gauged. 
    In particular, given a Lax connection in standard 4d CS, we expect the associated model to have a `gauged' version that can be found from the doubled theory.
    This is under the proviso that all boundary conditions of $A$ can be replaced by gauged versions \`a la the gauged chiral boundary conditions etc. discussed above. 

    Once again, for simplicity's sake, we fix $\Sigma = \mathbb{R}^2$, and work with the light-cone coordinates $x^{+}$, and $x^{-}$. 
    The metric and volume forms will be $\eta^{+-} = 2, \eta^{++} = \eta^{--} = 0$ and $d^2 x = dx^{+} \wedge dx^{-}$, as before. \

    \begin{table}[h!]
        \centering
        \def\arraystretch{1.5}
        \begin{tabular}{c|c|c}
             $\omega$ & \multicolumn{2}{|c}{Gauged P-defects} \\ 
            \hline
             \multirow{3}{*}{$k \dfrac{dz}{z}$} & Type & Position in $\mathbb{C}P^1 $ \\ 
             \cline{2-3}
             & Gauged chiral & $0$  \\
             \cline{2-3}
             & Gauged anti-chiral & $\infty$
        \end{tabular} 
        \caption{The defect configuration in example V.}
        \label{table:7}
    \end{table} 

    Consider the defect configuration given in Table \ref{table:7}. 
    As there are no Z-defects in this set up, it follows that $\mathcal A$ and $\mathcal B$ are given by the expressions: 
    \begin{equation}
        \mathcal A = \mathcal A^c_+ dx^+ + \mathcal A^c_- dx^{-} \ca \qquad \mathcal B = \mathcal B^c_+ dx^+ + \mathcal B^c_- dx^{-}
    \end{equation}
    Hence, when the genuine boundary conditions are substituted into equation \eqref{Lax from bc A}, we find the (trivial) simultaneous equations: 
    \begin{equation}
        \mathcal A_{-}^c = g^{-1} \mathcal B_{-}^c g +g^{-1} \partial_{-} g \ca \qquad \mathcal A^c_+ = \mathcal B^c_{+}
    \end{equation} 
    in which we've used $g_0 = g$ and $g_{\infty} = 1$. 
	Upon substituting these equations into the unified gauged $\sigma$-model, we find the gauged WZW model: 
    \begin{equation}
        S_{\text{GWZW}} (g,\mathcal{B}^c) = S_{\text{WZW}}(g) + \frac{k}{2 \pi} \int_{\mathbb{R}^2} d^2 x \left( \tr{\partial_+ g g^{-1}}{\mathcal B_-^c} - \tr{g^{-1} \partial_- g}{\mathcal B_+^c} - \tr{g \mathcal{B}_{+} g^{-1}}{\mathcal B_{-}^c} + \tr{\mathcal B_+}{\mathcal B_-} \right) \ca
    \end{equation}  
    whose equations of motion are known to be the flatness of the Lax connections $\mathcal A$ and $\mathcal B$: 
    \begin{equation}
        \partial_+ (g^{-1} \mathcal B_{-}^c g +g^{-1} \partial_{-} g) - \partial_{-} \mathcal B_+^c +[\mathcal B_+^c , g^{-1} \mathcal B_{-}^c g +g^{-1} \partial_{-} g] = 0 \ca \qquad \partial_+ \mathcal B_-^c - \partial_- \mathcal B_+^c + [\mathcal B_+^c , \mathcal B_-^c] = 0
    \end{equation}
    together with the gauging constraints given by substituting the Lax connections into \eqref{eq: gauging constraints in terms of Laxes}: 
    \begin{equation}
        (g \partial_+ g^{-1} + g \mathcal B_+^c g^{-1}) \proh = \mathcal B_{+}^c \ca \qquad (g^{-1} \partial_- g + g^{-1} \mathcal B_-^c g) \proh = \mathcal B_{-}^c \fp  
    \end{equation} 
	
	\subsection{The Nilpotent Gauged WZW Model}
	
	\label{section:nilpotent gwzw} 

    So far in our work, we have assumed $H$ is chosen to ensure $G/H$ is a reductive homogeneous space. 
    The purpose of this section is to illustrate that similar constructions are possible even if $H$ does not satisfy this requirement. 
    Our goal is the construction of the nilpotent gauged WZW model introduced in \cite{Forgacs:1989ac,Balog:1990mu}, from which the conformal Toda field theories and W-algebras can be found. 

    The traditional WZW model has the symmetry group, $G_{L} \times G_{R}$ where $G_{L}$ (resp. $G_R$) acts from the left $g \rightarrow u(x^+)g$ (resp. right $g \rightarrow g \bar{u}(x^-)$). 
    What differentiates the nilpotent gauging from the more traditional version is that these two symmetries can be gauged independently from each other. 
    This produces a $\sigma$-model whose target space is $G/(N^{-} \times N^{+})$. 
    To do this, we choose two maximal nilpotent subgroups, $N^+$ and $N^-$, of $G$ that are associated with the positive and negative roots, respectively. 
    Their Lie algebras are $\mathfrak n^+$ and $\mathfrak n^-$. 
    Next, we introduce a pair of gauge fields, $C \in \mathfrak n^+$ and $B \in \mathfrak n^-$, which, once coupled to the model, gauge left and right symmetries by $N^+$ and $N^-$. 
    The Toda theories are recovered once we fix the gauge $C_{-} = B_{+} = 0$ and perform a Gauss decomposition, as discussed in \cite{Balog:1990mu}. 
    
    In this section, we will assume $G = SL(N,\mathbb{C})$ and thus work in a basis where $\mathfrak{n}^{+}$ is the set of strictly upper triangular matrices, and $\mathfrak{n}^{-}$ is the set of strictly lower triangular matrices. 
    The case of general $G$ is recovered by replacing $\mathfrak{n}^{+}$ and $\mathfrak{n}^{-}$ by a pair of maximal nilpotent subalgebras.

	Consider a tripled version of the 4d CS model on $\mathbb{R}^2 \times \mathbb{C} P^{1}$ with three gauge fields $A \in \mathfrak{sl}(n)$, $B \in \mathfrak{n}^{-}$, $C \in \mathfrak{n}^{+}$:
	\begin{align}
		S_{\text{Tripled}} (A,B,C) = S_{\text{4dCS}}(A) - S_{\text{4dCS}}(B) &- S_{\text{4dCS}}(C) \label{nilpotent tripled} \\
		+ \frac{1}{4 \pi} \int_{\mathbb{R}^{2}_{0}} ( \tr{A}{C} + 2 \tr{A_{-}dx^{-}}{ \mu dx^{+}} )
		&- \frac{1}{4 \pi} \int_{\mathbb{R}^2_{\infty}} ( \tr{A}{ B} + 2 \tr{A_{+} dx^+ }{\nu dx^{-}}  ) \ca \nonumber
	\end{align}
	where:
	\begin{equation}
		\omega = \frac{dz}{z} \ca \label{6.2}
	\end{equation}
 	and in which $\mu \in \mathfrak{n}^{-}$ and $\nu \in \mathfrak{n}^{+}$ are constants. 
    We take $A,B$ and $C$ to be in the adjoint representation of $\mathfrak{g}$. 

    The Lie algebra $\mathfrak{sl}(N)$ can be decomposed into three subalgebras via $\mathfrak{sl}(N) = \mathfrak{n}^+ \oplus \mathfrak{t} \oplus \mathfrak{n}^-$, where $\mathfrak{t}$ is a Cartan subalgebra. 
    We denote the bases of each of the Lie algebras $\mathfrak n^+$, $\mathfrak{n^-}$ and $\mathfrak{t}$ in this decomposition by $\{e_{\alpha}\}$, $\{e_{-\beta}\}$, and $\{h_{\gamma}\}$, respectively. 
    The index used to label each basis element is a root of $\mathfrak{sl}$, whose space we denoted by $\Phi$. 
    This basis is chosen such that it satisfies: 
	\begin{equation}
		\tr{e_{\alpha}}{e_{\beta}} = \frac{2}{\alpha^2} \delta_{\alpha,-\beta} \ca \qquad \tr{h_{\gamma}}{h_{\tau}} = \gamma^{\vee} \cdot \tau^{\vee} \ca \qquad 	\tr{e_{\alpha}}{h_{\gamma}} = 0 \ca \label{7.64}
	\end{equation}
	where $\gamma, \tau \in \Delta$, $\alpha, \beta \in \Phi$, and $\alpha^{\vee} = 2 \alpha / \alpha^{2}$ is the coroot \cite{kirillov_jr_2008,Goddard:1986bp}. 
    See Appendix \ref{Cartan-Weyl Basis} for further details.  
    If we expand the actions $S_{\text{4dCS}}(B)$ and $S_{\text{4dCS}}(C)$ into their Lie algebra components, it is clear that $S_{\text{4dCS}}(B)=S_{\text{4dCS}}(C)=0$ by the first equation in \eqref{7.64}. 
    Hence the action \eqref{nilpotent tripled} reduces to:
	\begin{gather}
		S_{\text{Tripled}} (A,B,C) = S_{\text{4dCS}}(A) + \frac{1}{4 \pi} \int_{\mathbb{R}^{2}_{0}} ( \tr{A}{C} + 2 \tr{A_{-}dx^{-}}{ \mu dx^{+}} )
		\\ -\frac{1}{4 \pi}  \int_{\mathbb{R}^2_{\infty}} ( \tr{A}{ B} + 2 \tr{ A_{+} dx^+ }{\nu dx^{-}}  ) \ca \label{tripled nilpotent reduced}
	\end{gather}
	where the fields $B$ and $C$ behave as Lagrange multipliers. 
    It thus follows that we have one bulk equation of motion:
	\begin{equation}
		\omega \wedge F(A) = 0 \ca
	\end{equation}
	where $A$ is gauge equivalent to a Lax connection $\lax{A}$ by $A = \hat{g} d \hat{g}^{-1} + \hat{g} \lax{A} \hat{g}^{-1}$. We note that as above $\hat{g}$ is defined by $A_{\bar{z}} = \hat{g} \partial_{\bar{z}} \hat{g}^{-1}$. 

	If we vary $A$, $B$ and $C$ together while using \eqref{integral over poles} and \eqref{generic residue integral} we find the defect equations of motion:
	\begin{gather}
		\int_{\mathbb{R}^2_{0}} \left( \tr{A-C}{\delta A} + \tr{A}{\delta C} + 2 \tr{ \delta A_{-} dx^{-}}{ \mu dx^{+}} \right) = 0 \ca	\label{6.66} \\
		\int_{\mathbb{R}^{2}_{\infty}} \left( \tr{A-B}{\delta A} + \tr{A}{ \delta B} + 2 \tr{\delta A_{+} dx^{+}}{ \nu dx^{-}} \right) = 0 \fp \label{6.67}
	\end{gather}
	We solve \eqref{6.66} by expanding the Lie algebra components into the subset $\mathfrak{t},\mathfrak{n}^{+}$ and $\mathfrak{n}^{-}$, and introduce nilpotent versions of gauged chiral boundary conditions:
    \begin{gather}
		A_{-}^{\mathfrak{n}^{+}}(0) = C_{-} \ca \quad A_{-}^{\mathfrak{n}^{-} \oplus \mathfrak{t}}(0) = 0 \ca \quad A_{+}^{\mathfrak{n}^{-}}(0) = \mu \fp \label{7.69} 
    \end{gather}
    where $A_{+}^{\mathfrak{n}^{+}}(0) = \mu$ implies $\delta A_{+}^{\mathfrak{n}^{+}}(0) = 0$.
    Similarly, \eqref{6.67} is solved by a nilpotent version of gauged anti-chiral boundary conditions:
	\begin{gather}
        A_{+}^{\mathfrak{n}^{-}}(\infty) = B_{+} \ca \quad A_{+}^{\mathfrak{n}^{+} \oplus \mathfrak{t}}(\infty) = 0 \ca \quad A_{-}^{\mathfrak{n}^{+}}(\infty) = \nu \ca \label{7.70}  
	\end{gather}

    Just as we have done throughout this work, we can use the archipelago conditions to reduce the action down to a two-dimensional one on the defects at $\omega$'s poles. 
    For the purposes of this analysis, it is enough to note that nilpotent boundary conditions are preserved by gauge transformations of $A$ which satisfy $u=1$ at $z=0$ and $z=\infty$.
    To show the archipelago gauge is accessible, we consider the gauge transformation generated by $u = \hat{g} \tilde{g}^{-1}$, where $\tilde{g}$ satisfies the archipelago conditions.
    By the third condition, we have $u = 1$ at $z=0$ and $z=\infty$.
    Thus, our boundary conditions are preserved.

	Hence, upon using the archipelago conditions \eqref{tripled nilpotent reduced} becomes:
	\begin{gather}
		S_{\text{Tripled}}(A,B,C) = S_{\text{USM}}(\tilde{g},\lax{A}) 
  + \frac{1}{4 \pi}  \int_{\mathbb{R}^{2}_{0}} ( \tr{A}{C} + 2 \tr{A_{-}dx^{-}}{ \mu dx^{+}} )
		\\ -\frac{1}{4 \pi} \int_{\mathbb{R}^2_{\infty}} ( \tr{A}{ B} + 2 \tr{ A_{+} dx^+ }{\nu dx^{-}}  ) \label{6.75}
	\end{gather}
	where $S_{\text{USM}}(\tilde{g},\lax{A})$ is the unified sigma model \eqref{unified sigma model action}.

    By demanding that $A$ be regular, our Lax connection is of the form:
	\begin{equation}
		\lax{A} = \lax{A}_{+}^{c} dx^{+} + \lax{A}_{-}^{c} dx^{-} \fp
	\end{equation}

    In the setup just described, the genuine boundary conditions used to determine the Lax connection are: 
    \begin{equation}
        A_{-} (0) = C_- \ca \qquad A_+ (\infty) = B_+ \fp
    \end{equation} 
    If we plug:
	\begin{equation}
		A_{i}\pp = g_{q} \partial_{i} g^{-1}_{q} + g_{q} \lax{A}_{i} g^{-1}_{q} \ca \label{6.13}
	\end{equation} 
    for $i= \pm$ into the above two expressions and fix $g_0 = g$ and $g_\infty = 1$, as earlier, we find the Lax connection is: 
    	\begin{equation}
		\lax{A} = B_{+} dx^{+} + \left(g^{-1} \partial_{-} g + g^{-1} C_{-} g \right) \fp \label{nilpotent lax}
	\end{equation} 
    If we substitute this into \eqref{6.75} we find the action for the nilpotent gauged WZW model:
	\begin{align}
		S_{\text{NGWZW}}(g,B_{+},C_{-}) = S_{\text{WZW}}(g) + \frac{k}{2 \pi} \int_{\mathbb{R}^{2}} d^2 x \, \left( \tr{\partial_{+}gg^{-1}}{C_{-}} \right. &- \tr{B_{+}}{ g^{-1}\partial_{-}g g^{-1} C_{-} g} \\
		&+ \left. \tr{\mu}{ C_{-}} + \tr{\nu}{ B_{+}} \right) \ca \nn
	\end{align}
	where $S_{\text{WZW}}(g)$ is the WZW model. 
    The equations of motion for the above action are indeed the flatness of the Lax connection \eqref{nilpotent lax} is flat, and the constraints: 
    \begin{align}
		(g \partial_{+} g^{-1} + g B_{+} g^{-1})|^{\mathfrak{n}^{-}} &= \mu \ca \label{6.86} \\
		(g^{-1} \partial_{-} g + g^{-1} C_{-} g)|^{\mathfrak{n}^{+}} &= \nu \fp \label{6.85}
	\end{align}
	found from the leftover boundary conditions. 
 
	\section{Conclusion}

	\label{conclusion}

	We have reviewed the recent work of Costello and Yamazaki \cite{Costello:2019tri}, and Delduc et al. \cite{Delduc:2019whp}. In these papers, it was shown that one could solve the equations of motion of 4d CS theory (with two-dimensional defects inserted into the bulk) by defining a class of group elements $\{\hat{g}\}$ in terms of $A_{\bar{z}}$. Given a solution to the equations of motion, one finds an integrable sigma model by substituting the solution back into the 4d CS action. These sigma models are classical field theories on the defects inserted into the 4d CS theory. In \cite{Delduc:2019whp} it was shown the equivalence class of Lax connections, $\lax{A}$, of an integrable sigma model are the gauge invariant content of $A$, where $\lax{A}$ is found from $A$ by performing the Lax gauge transformation \eqref{eq: A in the archipelago gauge}. That $\lax{A}$ satisfies the conditions of a Lax connection was due to the Wilson lines and bulk equations of motion of $A$.

	In section \ref{doubled 4dcs section} we introduced the doubled 4d CS theory, inspired by an analogous construction in three-dimensional Chern-Simons \cite{Moore:1989yh}. In this section, we coupled together two 4d CS theory fields, where the second field was valued in a subgroup of the first, by introducing a boundary term. This boundary term had the effect of modifying the defect equations of motion enabling the introduction of new classes of gauged defects associated with the poles of $\omega$. In the rest of this section it was shown that the properties of 4d CS theory, such as its semi-topological nature or the unusual gauge transformation, are also present in the doubled 4d CS theory, even with the introduction of the boundary term.

	In section \ref{section:UGSM} we used the techniques of Delduc et al. in \cite{Delduc:2019whp} to derive the unified gauged sigma model action \eqref{unified sigma model action}. It was found that this model is associated to two Lax connections, one each for $A$ and $B$, and some boundary conditions associated with the defects inserted in the bulk of the doubled theory. The unified gauged sigma model's equations of motion are the flatness of the Lax connections and the boundary conditions associated to the defects. We concluded in section \ref{section:nilpotent gwzw} by deriving the Gauged WZW and Nilpotent Gauged WZW models, from which one finds the conformal Toda field theories \cite{Balog:1990mu}.

	Before we finish, we wish to make some additional comments. The first of these is on the relation between the doubled four-dimensional action \eqref{doubled action} and its equivalent in three-dimensions:
	\begin{equation}
		S(A,B) = S_{\text{CS}}(A) - S_{\text{CS}}(B) - \frac{k}{4 \pi} \int_{M} d \tr{A}{B} \label{3d doubled}
	\end{equation}
	In \cite{Yamazaki:2019prm} it was proven that the 4d CS action for $\omega = dz/z$ is $T$-dual to the three-dimensional Chern-Simons action. By Yamazaki's arguments it is clear that the boundary term of the doubled action \eqref{doubled action} for $\omega = dz/z$ is $T$-dual to the boundary term of \eqref{3d doubled}, hence \eqref{doubled action} and \eqref{3d doubled} are $T$-dual. As a result, we expect that arguments analogous to those used in section \ref{section:UGSM} can be used to derive the gauged WZW model from \eqref{3d doubled}. It is important to note that this is different to the derivation of the gauged WZW model from Chern-Simons theory given in \cite{Moore:1989yh}. This is because the introduction of the boundary term leads to a modification of the defect equations of motion and therefore the boundary conditions. This contrasts with the construction given in \cite{Moore:1989yh} where a Lagrange multiplier was used to impose the relevant boundary conditions.


    Our second comment concerns the boundary conditions defined above. 
    The boundary conditions used in the doubled theory are not exhaustive; just as the two chiral, and Dirichlet conditions are not exhaustive in the standard theory.
    In \cite{Costello:2017dso,Costello:2019tri} the authors solve the defect equations of motion by requiring that $A$ be valued in an isotropic subalgebra $\mathfrak{l}$ at simple poles of $\omega$ -- here
    \textit{Isotropic} means that for $a,b \in \mathfrak{l}$ we have $\tr{a}{b}=0$.
    Similarly, in the doubled theory one can introduce a gauged version of this isotropic boundary condition.
    Suppose there exists an isotropic subalgebra of $\mathfrak{g}$ in $\mathfrak{f}$, the gauged isotropic condition is then defined by the requirement that $A_{\pm}^\mathfrak{h} = B_{\pm}$ and $A_{\pm}^{\mathfrak{f}} \in \mathfrak{l}$.
    This opens up the possibility for a $T$-duality between certain gauged models, as we now argue.

    In \cite{Delduc:2019whp} a somewhat more restricted version of both 4d CS and the isotropic boundary conditions were considered.
    In particular, reality conditions were imposed such the unified sigma model action was real.
    This requirement meant that first order poles of $\omega$ must be considered in pairs, $(q_{\pm})$, such that they are either: (a) on the real line or (b) complex conjugates.
    In case (a) one introduces a Manin pair $(\mathfrak{d},\mathfrak{l})$, where\footnote{$\mathfrak{g}^{\mathbb{R}}$ is a real form of $\mathfrak{g}$.} $\mathfrak{d} = \mathfrak{g}^{\mathbb{R}} \oplus \mathfrak{g}^{\mathbb{R}}$ has a subalgebra $\mathfrak{l}$ which is maximally isotropic (or \textit{lagrangian}) with respect to the bilinear form $\left\langle \! \! \tr{(A|_{q_{+}},A|_{q_{-}})}{ \delta (A|_{q_{+}},A|_{q_{-}})} \! \! \right\rangle \vcentcolon = \epsilon^{ij} (\eta_{q_{+}}^{0} \tr{A_{i}|_{q_{+}}}{\delta A_j|_{q_{+}}} + \eta_{q_{-}}^{0} \tr{A_{i}|_{q_{-}}}{\delta A_j|_{q_{-}}})$ on $\mathfrak{d}$.
    The restricted isotropic boundary condition of \cite{Delduc:2019whp} is then defined by requiring that $(A_{\pm}|_{q_{+}},A_{\pm}|_{q_{-}}) \in \mathfrak{l}$.
    In case (b) one does the same thing, but with $\mathfrak{d}$ replaced by $\mathfrak{g}$.
    We expect a similar, gauged, construction to exist in the doubled theory.


    In \cite{Delduc:2019whp} the Manin double construction is extended to a Manin triple $(\mathfrak{d},\mathfrak{l}_1,\mathfrak{l}_2)$ - where $\mathfrak{l}_{1}$ and $\mathfrak{l}_{2}$ are both isotropic subalgebras of $\mathfrak{d}$ such that\footnote{Here $\plusdot$ denotes the direct sum as a vector space.} $\mathfrak{d} = \mathfrak{l}_{1} \plusdot \mathfrak{l}_{2}$.
    Given the Manin triple, one solves the defect equations of motion by imposing that $A$ is valued in $\mathfrak{l}_1$ (or $\mathfrak{l}_2$) at both poles.
    It was suggested that for a fixed $\omega$ the models found by imposing Manin triple boundary conditions in case (a) should be Poisson-Lie $T$-dual to those found from case (b), where one has also imposed Manin triple boundary conditions. 
    We hope the same is true in a gauged Manin triple boundary condition.

    Our third comment concerns the scalars $\Delta_q$.
    In the example of GWZW + BF theory we saw that $\Delta$ deforms the equations of motion of the GWZW model realised when $\Delta = 0$.
    We expect this to be a general feature of such models, with $\Delta_{q}$ inducing integrable deformations of a theory obtained when $\Delta_q = 0$, \`a la the conformal Toda models found from deformations of Toda theories \cite{conformalaffinetoda,Constantinidis:1992hs,Papadopoulos:1994iw}.
    We intend to explore this elsewhere.

	Finally, we hope that one can find other new integrable gauged sigma models using the construction defined in section \ref{section:UGSM}. This being said, there are several other problems which we have not discussed in this paper, but which we plan to cover in the future. These include $\lambda$- \cite{Hollowood:2014rla,Sfetsos:2013wia}, $\eta$- \cite{Klimcik:2002zj,Delduc:2013fga}, and $\beta$-deformations \cite{Lunin:2005jy,Kawaguchi:2014qwa,Osten:2016dvf}, this is expected to be similar to \cite{Chen:2021qto} and \cite{Fukushima:2020kta,Fukushima:2020dcp,Fukushima:2021eni}; the generation of affine Toda models from 4d CS theory; the generation of gauged sigma models associated to a higher genus choice of $C$ -- we expect this to be analogous to the discussion near the end of \cite{Costello:2019tri}; how to find a set of Poisson commuting charges from $\lax{A}$ and $\lax{B}$ such that $\lax{A}$ and $\lax{B}$ are Lax connections; related to this is, the connection between our construction of gauged sigma models and that given by Gaudin models, this is likely similar to \cite{Vicedo:2019dej}; the quantum theory of the doubled action; and finally whether the results of \cite{Bittleston:2020hfv} can be repeated for the doubled action, enabling us to find higher dimensional integrable gauged sigma models.

	\appendix

	\section{The K{\"u}nneth Theorem and Cohomology}
	\label{appendix:DeRham}
	The K{\"u}nneth theorem gives one a relation between the cohomologies of a product space and the cohomologies of the manifolds which it is constructed from:
	\begin{equation}
		H^k ( X \times Y ) = \bigoplus_{i + j =k} H^{i}(X) \otimes H^{j}(Y) \fp
	\end{equation}
	The de Rham cohomology for $\mathbb{R}^n$ is:
	\begin{equation}
		H^{k}( \mathbb{R}^n ) \cong \begin{cases}
		\mathbb{R} \ca \; & \text{if} \; k=0, \\
		0 \ca \; & \text{otherwise.}
		\end{cases}
	\end{equation}
		While for $\mathbb{C} P^{n}$ this is:
	\begin{equation}
		H^{k} (\mathbb{C} P^{n} ) \cong \begin{cases}
		\mathbb{R} \ca \; & \text{for k even and } 0 \leq k \leq 2n, \\
		 0\ca \; & \text{otherwise.}
		\end{cases} 
	\end{equation}

	\section{Conventions for the WZW and Gauged WZW Models}

	\label{WZW convetions}

	The WZW model is constructed from the field $g : \mathbb{R}^{2} \rightarrow G$, where $G$ is a complex Lie group, and is defined by the action:
	\begin{equation}
		S_{\text{WZW}}(g) = \frac{k}{8 \pi} \int_{\mathbb{R}^2} d^2 x \sqrt{-\eta} \eta^{\mu \nu} \, \tr{g^{-1} \partial_{\mu} g }{g^{-1} \partial_{\nu} g} + \frac{k}{12 \pi} \int_{B} \tr{g^{-1} d g}{g^{-1} d g \wedge g^{-1} d g}\ca \label{wzwappendix}
	\end{equation}
	where $d^2x = dx^{+} \wedge dx^{-}$, $\eta^{\mu \nu}$ a metric on $\mathbb{R}^{2}$, $\eta$ the determinant of $\eta_{\mu \nu}$, and $\hat{g}$ the extension of $g$ into the three-dimensional manifold $B$, where $\partial B = \mathbb{R}^{2}$. In this paper we take $B = \mathbb{R}^{2} \times [0,R_{0}]$ with light-cone coordinates $x^{\pm}$ on $\mathbb{R}^2$ and metric $\eta^{+-} = 2, \eta^{++} = \eta^{--} = 0$. Our light-cone coordinates are connected to the Lorentzian coordinates $x^{0},x^{1}$ by $x^{+} = x^{0} + x^{1}$ and $x^{-} = x^{0} - x^{1}$ with the Minkowski metric $\eta_{00} = -\eta_{11} = 1, \eta_{01} = 0$.

	The WZW action is invariant under transformations of the form $g \rightarrow u(x^{+})g\bar{u}(x^{-})^{-1}$ in $G_{L} \times G_{R}$ where $u \in G_{L}$ and $\bar{u} \in G_{R}$. To show this invariance one defines an extension of $u$ and $\bar{u}$ into $B$, denoted $\hat{u}$, and uses the Polyakov-Wigmann identity:
	\begin{equation}
		S_{\text{WZW}}(gh) = S(g) + S(h) + \frac{k}{2 \pi} \int_{\mathbb{R}^2} dx^{+} \wedge dx^{-} \tr{ g^{-1} \partial_{-} g}{ \partial_{+} h h^{-1}} \label{PW identity} \ca
	\end{equation}
	to expand $S_{\text{WZW}}(ug\bar{u})$ into a sum over WZW terms. Upon doing this one finds all terms other than $S_{\text{WZW}}(g)$ vanish. On $B = \mathbb{R}^{2} \times [0,R_{0}]$ we parametrise $[0,R_{0}]$ by $z$ and define the extension $\hat{u}$ such that $\hat{u}|_{z=0} = \bar{u}$ and $\hat{u}|_{z=R_{0}} = u$, this ensures a cancellation of the Wess-Zumino terms associated to $u$ and $\bar{u}$. All other terms vanish due to $\partial_{-}u = \partial_{+}\bar{u} = 0$. 

	From the variation $g \rightarrow g + \delta g$ in \eqref{wzwappendix} one finds the variation of the action:
	\begin{equation}
		\delta S(g) = - \frac{k}{2 \pi} \int_{\mathbb{R}^2} dx^{+} \wedge dx^{-} \tr{g^{-1} \delta g}{ \partial_{+}(g^{-1}\partial_{-}g)} = - \frac{k}{2 \pi} \int_{\mathbb{R}^{2}} dx^{+} \wedge dx^{-} \tr{\delta g g^{-1}}{ \partial_{-}(\partial_{+} g g^{-1})} \ca
	\end{equation}
	which gives the equations of motion:
	\begin{equation}
		\partial_{+}(g^{-1}\partial_{-}g) = \partial_{-}(\partial_{+}gg^{-1}) = 0 \ca
	\end{equation}
	where $J_{+} = \partial_{+} g g^{-1}$ and $J_{-} = g^{-1} \partial_{-} g$ are the currents of the model. These equations have the solution:
	\begin{equation}
		g(x^{+},x^{-}) = g_{l}(x^{+}) g_{r}(x^{-})^{-1} \ca
	\end{equation}
	where $g_{l}$ ($g_{r}$) is a generic holomorphic (anti-holomorphic) map into $G$.

	One can define a version of the WZW model where the symmetry $g \rightarrow ug\bar{u}^{-1}$ is gauged by a group $H \subseteq G$, this gives an action to the coset models \cite{Goddard:1984hg,Goddard:1984vk,Goddard:1986ee} as shown in \cite{Karabali:1989dk,Karabali:1988au,Hwang:1993nc,Gawedzki:1988hq,Gawedzki:1988nj}. This gauged WZW model can be found from the normal WZW model by applying the Polyakov-Wigmann identity \eqref{PW identity} to:
	\begin{equation}
		S_{\text{Gauged}}(g,h,\tilde{h}) = S_{\text{WZW}}(hg\tilde{h}^{-1}) - S_{\text{WZW}}(h \tilde{h}^{-1}) \label{B6} \ca
	\end{equation}
	where $h(x^{+},x^{-}), \tilde{h}(x^{+},x^{-}) \in H$. It is clear that this equation is invariant under the transformation $g \rightarrow ugu^{-1}, h \rightarrow hu^{-1}, \tilde{h} \rightarrow \tilde{h}u^{-1}$ for $u(x^{+},x^{-}) \in H$. After expanding \eqref{B6} and setting $B_{-} = h^{-1} \partial_{-} h$ and $B_{+} = \tilde{h}^{-1} \partial_{+} \tilde{h}$ one finds gauged WZW model action:
	\begin{align}
		S_{\text{Gauged}}(g,B_{+},B_{-}) = S_{\text{WZW}}(g) + \frac{k}{2 \pi} \int_{\mathbb{R}^{2}} dx^{+} \wedge dx^{-} & \left( \tr{\partial_{+}g g^{-1}}{B_{-}} \right. \\
		&- \left. \tr{B_{+}}{ g^{-1} \partial_{-}g} - \tr{gB_{+}g^{-1}}{B_{-}} + \tr{B_{+}}{B_{-}}\right) \ca \nn
	\end{align}
	where the symmetry $g \rightarrow ugu^{-1}, h \rightarrow hu^{-1}, \tilde{h} \rightarrow \tilde{h}u^{-1}$ corresponds to the gauge transformation:
	\begin{align}
		g \rightarrow u g u^{-1} \ca \qquad B_{\pm} \rightarrow u(\partial_{\pm} + B_{\pm})u^{-1} \ca 
	\end{align}
	for $u(x^{+},x^{-}) \in H$. This gauge symmetry means the orbits of $G$ which are mapped to each other by the action of $H$ are identified and therefore physical equivalent, hence the target space of the gauged WZW model is the coset $G/H$.

	It is important to note that two conventions for the  WZW model and Polyakov-Wigmann identity exist which are related by $g \rightarrow g^{-1}$, $h \rightarrow h^{-1}$. Further still, four conventions for the gauged WZW models exist found by taking $g \rightarrow g^{-1}$ and $B_{+} \rightarrow -B_{+}$ in various combinations.

	\section{The Cartan-Weyl Basis}

	\label{Cartan-Weyl Basis}

Here we collect some facts about Lie algebras from \cite{Fuchs:1997jv}.
A semi-simple Lie algebra $\mathfrak{g}$ can be decomposed into three subalgebras 
$\mathfrak{g} =  \mathfrak{t} \oplus \mathfrak{n}^+ \oplus \mathfrak{n}^{-}$. The first subalgebra is a Cartan subalgebra $\mathfrak{t}$ which is a maximal set of commuting semi-simple\footnote{An element $x \in \mathfrak{g}$ is semi-simple if the matrix of eigenvalues formed by the adjoint action $\text{ad}_x$ is diagonalisable.} elements of $\mathfrak{g}$. 
We take $\{H_i\}$ to be a basis of $\mathfrak{t}$.
We can choose a basis of $\mathfrak{n}^+$ to be $\{e_\alpha\}$ where the set $\{\alpha\}=\Phi^+$ is called the set of positive roots. Similarly, $\{e_{-\alpha}\}$ span $\mathfrak{n}^-$ and $\{-\alpha\} = \Phi^-$ is the set of negative roots.

The Killing form on $\mathfrak{g}$ is $K(x,y) = \text{Tr}(\text{ad}_{x} \circ \text{ad}_{y})$, where $\text{ad}_x$ denotes the adjoint action of $x$, $\circ$ the composition of maps, and $\text{Tr}$ over linear maps. 
Let $\tr{\cdot}{\cdot}$ denote a symmetric invariant bilinear form proportional to the Killing form.
We can always choose the basis elements $H_i$ to be orthonormal.
With these choices, we can take the commutators to be
%
%
%
	\begin{align}
		[H_{i},H_{j}] &= 0 \ca & [H_{i},e_{\pm \alpha}] &= \pm \alpha^{i} e_{\pm \alpha} \ca \label{D1} \\
		[e_{\alpha},e_{-\alpha}] &= \frac{2 \alpha^{i}}{\alpha^{2}} H_{i} \ca & [e_{\pm \alpha},e_{\pm \beta}] &= \epsilon(\pm \alpha, \pm \beta) e_{\pm \alpha \pm \beta} \fp \label{D2}
	\end{align}
	where $\epsilon(\pm \alpha, \pm \beta)$ is a structure constant for any pair of $\pm$, $\alpha^{i}$ the $i$-th element of $\alpha \in \Phi^+$, and $\alpha^2 = \tr{\alpha}{\alpha}$. If in the final equation $\pm \alpha \pm \beta \not\in \Phi^{\pm}$ then $\epsilon(\pm \alpha, \pm \beta) = 0$.

	Let $\Delta$ denote the set of generators of $\Phi^+$, which are called simple roots. For each root $\alpha \in \Phi$ one can define an element of the Cartan Subalgebra given by $h_{\alpha} = \alpha^{\vee}_{i} H_{i}$, where $\alpha^{\vee}_{i} = 2 \alpha_{i}/\alpha^2$ is the coroot. The set of elements $\{h_{\alpha}\}$ labelled by a simple root, $\alpha \in \Delta$, form a basis of the Cartan subalgebra. From this result the equations (\ref{D1},\ref{D2}) can be rewritten as:
	\begin{align}
		[h_{\gamma},h_{\tau}] &= 0 \ca & [h_{\gamma},e_{\pm \beta}] &= \pm \gamma^{\vee} \cdot \beta e_{\pm \beta} \ca \\
		[e_{\alpha},e_{-\alpha}] &= h_{\alpha} \ca & [e_{\pm \alpha},e_{\pm \beta}] &= \epsilon(\pm \alpha, \pm \beta) e_{\pm \alpha \pm \beta} \ca 
	\end{align}
	where $\gamma,\tau \in \Delta$ and $\alpha, \beta \in \Phi^{+}$. Note, to each root $\alpha \in \Phi^{+}$ we can associate an $\mathfrak{sl} (2)$ given by $\mathfrak{g}_{\alpha} = \{e_{\alpha},e_{-\alpha},h_{\alpha}\}$.

	
	The inner product $\tr{h_{\alpha}}{h_{\beta}}$ is found by noting the basis elements $\{H_{i}\}$ are orthonormal, i.e. $\tr{H_{i}}{H_{j}} = \delta_{ij}$, hence:
	\begin{equation}
		\tr{h_{\alpha}}{h_{\beta}} = \frac{4 \alpha_{i} \beta_{j}}{\alpha^2 \beta^2} \tr{H_{i}}{H_{j}} = \tr{\alpha^{\vee}}{ \beta^{\vee}} \ca
	\end{equation}
	where $\tr{\alpha^{\vee}}{\beta^{\vee}}$ is the symmetrised Cartan matrix. The final bilinear form to be found is $\tr{e_{\alpha}}{e_{-\alpha}}$. Using the identity $\tr{X}{[Y,Z]} = \tr{[X,Y]}{Z}$ it is clear that:
	\begin{equation}
		\tr{\alpha^{\vee}}{ \alpha }\, \tr{e_{\alpha}}{e_{-\alpha}} = \tr{h_{\alpha}}{[e_{\alpha},e_{-\alpha}]} = \tr{h_{\alpha}}{h_{\alpha}} = \frac{4}{\alpha^2} \ca
	\end{equation}
	hence our trace in the basis $\{h_{\gamma},e_{\alpha},e_{-\beta}\}$ is:
	\begin{equation}
		\tr{e_{\alpha}}{e_{\beta}} = \frac{2}{\alpha^2} \delta_{\alpha,-\beta} \ca \qquad \tr{h_{\gamma}}{h_{\tau}} = \gamma^{\vee} \cdot \tau^{\vee} \ca \qquad \tr{e_{\alpha}}{h_{\gamma}} = 0 \ca \label{D9}
	\end{equation}
	where $\gamma, \tau \in \Delta$ and $\alpha, \beta \in \Phi$.

	\printbibliography

\end{document}